\documentclass{article}

 \usepackage[preprint]{neurips_2025}

\makeatletter
\renewcommand{\@notice}{}
\makeatother

\usepackage[utf8]{inputenc} 
\usepackage[T1]{fontenc}    
\usepackage{hyperref}       
\usepackage{url}            
\usepackage{booktabs}       
\usepackage{amsfonts}       
\usepackage{nicefrac}       
\usepackage{microtype}      
\usepackage{xcolor}         
\definecolor{citeblue}{RGB}{0,51,170}     
\definecolor{refdarkred}{RGB}{153,0,0}    
\definecolor{urlteal}{RGB}{0,110,120}     
\hypersetup{colorlinks=true, citecolor=citeblue, linkcolor=refdarkred, urlcolor=urlteal, filecolor=urlteal} 
\usepackage{tikz}
\usetikzlibrary{arrows.meta,positioning,shapes.geometric,fit,backgrounds}
\usepackage{amsmath} 
\usepackage{amssymb}
\usepackage{array}
\usepackage{tabularx}
\usepackage{placeins}       

\newcommand{\equalcontrib}{\textsuperscript{\ensuremath{\dagger}}}
\newcommand{\correspon}{\textsuperscript{*}}
\title{Statopia: A Prover--Library Ecosystem for Automated Formalization and Proof Discovery in Statistics}

\author{%
\begin{tabular}{c}
\bfseries
Ziyuan Chen\textsuperscript{1}\equalcontrib
\quad
Tong Gan\textsuperscript{2}\equalcontrib
\quad
Wei Zhang\textsuperscript{3}\equalcontrib
\quad
Ruofei Fang\textsuperscript{4}
\\[2pt]
\bfseries
Boyue Tang\textsuperscript{1}
\quad
Zhenhua Lin\textsuperscript{2}\correspon
\quad
Fang Yao\textsuperscript{1}\correspon
\\[6pt]
\normalfont\small
\textsuperscript{1}School of Mathematical Sciences, Center for Statistical Science, Peking University
\\
\normalfont\small
\textsuperscript{2}Department of Statistics and Data Science, National University of Singapore
\\
\normalfont\small
\textsuperscript{3}Department of Statistics and Data Science, School of Management, Fudan University
\\
\normalfont\small
\textsuperscript{4}Department of Computer Science/Statistics, The University of British Columbia
\end{tabular}
}

\makeatletter
\renewcommand{\@thanks}{%
  \footnotetext[2]{Ziyuan Chen, Tong Gan and Wei Zhang are co-first authors.}%
  \footnotetext[1]{Zhenhua Lin and Fang Yao are co-corresponding authors. E-mails: \texttt{linz@nus.edu.sg} and \texttt{fyao@math.pku.edu.cn}.}%
}
\makeatother

\begin{document}

\maketitle
\begin{abstract}
Theoretical results in statistics and machine learning rest on long, technically delicate proofs, and fluent arguments generated by large language models are not necessarily correct. Automated formalization makes such arguments mechanically checkable, and automated proving goes further by discovering machine-checked proofs for statements whose proofs are incomplete or missing altogether. Both require reconstructing assumptions from source materials, accessing specialized mathematical infrastructure and managing long proof developments. We present \textbf{\emph{Statopia}}, an ecosystem that jointly develops \textbf{\emph{Statopia Prover}}, an independent agent specialized for formalizing and proving results in statistics and machine learning, and \textbf{\emph{Statlib}}, a Lean~4 library built with the prover and organized for automated use. The prover controls its own execution without relying on external coding agents. It is built for formal proof discovery and checking existing proofs. A natural-language proof, when available, guides its search but is not required: from a statement alone, the prover develops the argument itself and constructs a Lean proof. Its workflow integrates source preparation, statement review, library supported planning and proof construction. Dynamic proof dependency graphs and task groups support parallel, resumable proof development. Statlib's shared definitions and theorem interfaces are organized for retrieval and composition by agents, and its coverage of statistics keeps expanding. Verified results and usage feedback from the prover inform reviewed library updates, so that proof development drives this expansion. A cross-platform application provides an interactive workspace for uploading source files and running tasks locally or in the cloud. Case studies demonstrate proof discovery and checking for individual theorems, statistical paper developments and the transformation of verified proof structures into reusable library infrastructure.
\end{abstract}

\vspace{-0.5em}
\vspace{0.5em}

\newpage

\section{Introduction}
\label{sec:intro}

Theoretical results in statistics and machine learning, such as consistency, convergence rates, coverage guarantees and generalization bounds, often rest on long and technically delicate proofs. Errors in such proofs can survive peer review. For example, in a published analysis of the geometric ergodicity of Hamiltonian Monte Carlo \citep{livingstone2019geometric}, a typographical error in one condition left a gap in the proof of a corollary; \citet{gruffaz2026nuts} identified the gap, and the authors corrected it in a 2026 corrigendum. Large language models (LLMs) can now generate statistical arguments at scale, which makes the problem more pressing: a fluent natural-language proof may still contain subtle errors, and reviewing it with another model provides no guarantee of correctness. Formal verification offers a stronger standard: statements, assumptions and proofs are expressed in a formal language, and a proof assistant such as Lean checks the proofs against the encoded statements through a small trusted kernel. Because manual formalization requires substantial effort, automated formalization aims to make this standard accessible by translating mathematical sources into explicit formal statements and machine checkable proofs. Automated proving extends the same standard to statements that come with an incomplete proof or none at all, such as a lemma stated without proof, a step left to the reader or a claim put forward by a researcher or a language model. A prover that searches for the proof itself can then verify such a statement formally, with Lean checking whatever proof it finds. Formal methods thus serve proof discovery as well as proof checking. The resulting artifacts can be inspected for their assumptions and dependencies, rechecked after revision and reused in subsequent developments.

Statistical and machine learning results, however, place three additional demands on automated formalization and proving, beyond the general mathematical settings in which general-purpose systems (Section~\ref{sec:related-work}) have recently advanced \citep{rammal2026formalizing,milikic2026leanflow,zhang2026leanmarathon,zhang2026dreamprover}. First, faithful formalization requires reconstructing the statistical setting, including the data generating mechanism, estimator, distributional assumptions, regularity conditions and asymptotic quantifiers, which may be scattered throughout a source. Lean checks the encoded proposition, so successful compilation alone does not establish fidelity to the intended result \citep{zhang2026beyond,ammanamanchi2026faults}. Second, proofs often require domain level infrastructure, such as concentration inequalities, empirical process results and risk decompositions, beyond the general foundations provided by Mathlib \citep{mathlib2020}. Effective reuse depends on whether these results can be found and their interfaces composed. Third, statistical arguments combine probabilistic, analytic, optimization and matrix reasoning across long chains of dependencies. Where the source proof is incomplete or absent, a prover must construct these chains itself; it must also preserve verified progress, manage dynamic proof obligations and distinguish local Lean errors from problems in the mathematical plan or the available infrastructure.

Recent agent-based systems for statistics and machine learning have begun to address these demands through specialized formalization workflows, domain libraries and statement review. These systems typically run inside general-purpose coding agents such as Claude Code or Codex, so their agent loop, context management and tool execution come from a runtime built for general coding tasks (Section~\ref{sec:related-work} and Table~\ref{tab:ecosystem-comparison}). We introduce \emph{Statopia}, an ecosystem that jointly develops a native statistical proving system, \emph{Statopia Prover} (\url{https://prover.statopia.ai}), and a Lean~4 library for statistics and machine learning, \emph{Statlib} (\url{https://statlib.statopia.ai}). Statopia Prover is a full-stack prover: it implements its own agent loop, context management, proof state, tool execution and acceptance checks rather than running inside Claude Code, Codex or any other coding-agent harness, and its language model backends are replaceable. The central principle is reciprocal development: Statlib provides the mathematical vocabulary and reusable interfaces needed by the prover, while actual proof development provides evidence for improving the library. Statlib is itself developed with Statopia Prover, so the library grows from the same kind of verified proof developments that it supports. The aim is to reduce avoidable retrieval, interface adaptation and reconstruction work by designing the proving process and its supporting infrastructure together. Among the systems compared in Table~\ref{tab:ecosystem-comparison}, only Statopia combines such a self-contained proving runtime with a co-designed statistical library, proof-driven library evolution and a graphical workspace for local or cloud execution.

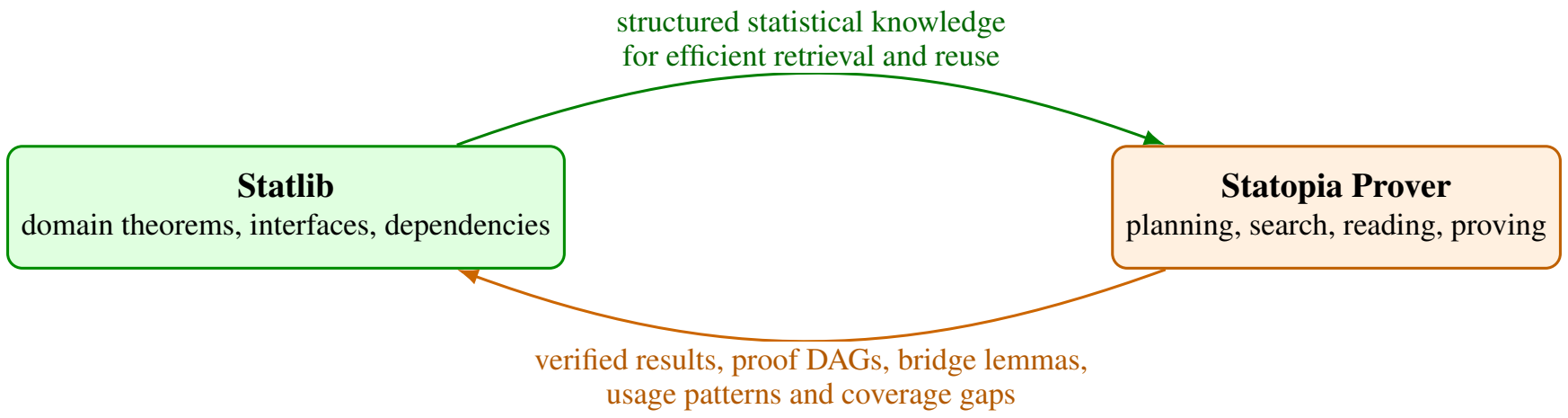
\begin{figure}[!ht]
\centering
\resizebox{0.82\linewidth}{!}{%
\begin{tikzpicture}[
    >={Latex[length=2.2mm]},
    node distance=5.6cm,
    every node/.style={align=center},
    box/.style={
        draw,
        rounded corners=1.5mm,
        thick,
        minimum width=3.3cm,
        minimum height=1.25cm,
        inner sep=4pt
    },
    librarybox/.style={
        box,
        draw=green!55!black,
        fill=green!12
    },
    proverbox/.style={
        box,
        draw=orange!75!black,
        fill=orange!12
    },
    librarylabel/.style={
        font=\footnotesize,
        align=center,
        text=green!40!black,
        fill=white,
        inner sep=2pt
    },
    proverlabel/.style={
        font=\footnotesize,
        align=center,
        text=orange!70!black,
        fill=white,
        inner sep=2pt
    }
]
\node[librarybox] (library) {
    \textbf{Statlib}\\
    {\footnotesize domain theorems, interfaces, dependencies}
};

\node[proverbox, right=of library] (prover) {
    \textbf{Statopia Prover}\\
    {\footnotesize planning, search, reading, proving}
};

\draw[->, thick, green!50!black, bend left=20]
    (library) to
    node[above, librarylabel]
    {structured statistical knowledge\\for efficient retrieval and reuse}
    (prover);

\draw[->, thick, orange!80!black, bend left=20]
    (prover) to
    node[below, proverlabel]
    {verified results, proof DAGs, bridge lemmas,\\usage patterns and coverage gaps}
    (library);
\end{tikzpicture}}
\caption{Prover and library co-design in the Statopia ecosystem. Statlib provides structured statistical knowledge that supports efficient retrieval and reuse during proof construction, while actual proving produces verified results and usage signals that guide subsequent library development.}
\label{fig:prover-library-loop}
\end{figure}

Statopia implements this design through a native agent specifically designed and optimized for formalizing and proving results in statistics and machine learning. The agent is built for formal proof discovery and checking. Its input can be a paper with its proofs, a theorem stated without proof or a bare Lean statement. When a proof is supplied, the prover completes it where needed and uses it as a guide; when none is supplied, it develops the mathematical argument itself. In either case, it searches for a Lean proof and introduces the intermediate lemmas that this proof requires, so a statement can be formally verified even when no natural-language proof is available. The prover supports source preparation and formalization review, checks retrieved declarations before reuse and coordinates larger developments through dynamic proof dependency graphs and task groups. These graphs allow the prover to work on many proof obligations at once: independent branches proceed in parallel, and a result can be developed against the Lean statement of a helper lemma while the lemma itself is still being proved, both within a task and across tasks; final acceptance still requires every supporting proof to be verified. Statlib supplies shared statistical definitions and organizes related declarations into coherent modules that expose main theorems and selected reusable intermediate results while keeping implementation-specific helpers private. Verified results, proof dependencies and recurring retrieval patterns then inform subsequent library refinement. An interactive workspace connects this process to source submission, review of formal targets, progress monitoring, resumable execution and inspection of the resulting Lean artifacts. Figure~\ref{fig:prover-library-loop} summarizes the interaction between the prover and the library.

Our contributions are summarized as follows.
\begin{enumerate}
    \item We develop \emph{Statopia Prover}, a full-stack agent specifically designed and optimized for formalizing and proving results in statistics and machine learning. It runs its own agent loop, proof state, tool execution and acceptance checks without relying on general-purpose coding agents such as Claude Code or Codex, and develops large proofs in parallel through dynamic proof dependency graphs.

    \item We introduce a \emph{co-designed prover and library ecosystem} for formal statistics and machine learning, connecting automated proof development with the growth of reusable statistical infrastructure. The library supports proof construction, while verified results and reuse needs identified by the prover guide subsequent library refinement.

    \item Using Statopia Prover, we develop \emph{Statlib}, a substantial Lean~4 library with broad statistical coverage, including high-dimensional, nonparametric statistics and beyond. It is organized for agent retrieval and reuse, and evolves through reviewed additions and interface refinements informed by proof development.

    \item We provide a \emph{cross-platform application} that integrates the prover and library within an interactive graphical workspace. Users can upload source files and carry out formalization tasks with either cloud or local execution.
\end{enumerate}

The remainder of this paper is organized as follows. Section~\ref{sec:related-work} reviews related work and compares Statopia with representative systems. Section~\ref{sec:overview} gives an overview of Statopia, from its use to its architecture and the key designs of its proof engine and library. Section~\ref{sec:statopia-prover} presents the formalization and proving workflow, from source preparation and statement review to proof discovery and construction, including the coordination of larger developments through dynamic proof dependency graphs and task groups. Section~\ref{sec:statlib} describes the mathematical content and organization of Statlib, together with the process for incorporating reusable results from proof development. Section~\ref{sec:evaluation} evaluates the ecosystem through statistical theorem tasks, guided mathematical problem solving and larger statistical formalization case studies. Section~\ref{sec:conclusion} concludes with a discussion of limitations and future directions.

\section{Related Work}
\label{sec:related-work}

Work related to Statopia follows two lines. General-purpose systems automate formalization and proving across mathematics, while work in statistics and machine learning develops specialized proving workflows and domain libraries. Table~\ref{tab:ecosystem-comparison} compares representative systems and libraries along six dimensions, grouping versions and companion works from the same project family into one row. Each column records whether at least one cited family member supports that capability; the comparison does not rank proof success or efficiency.

\begin{table}[!htbp]
\centering
\caption{Reported capabilities of Lean proving systems and domain libraries, grouped by project family. Marks are the column-wise union across the cited family members, not a claim that all capabilities coexist in one release. The domain-library column requires a project's own built-in or companion domain library; support for connecting external libraries alone does not qualify.}
\label{tab:ecosystem-comparison}
\footnotesize
\setlength{\tabcolsep}{2.5pt}
\renewcommand{\arraystretch}{1.08}
\providecommand{\sysname}[2]{#1\textsuperscript{#2}}
\begin{tabularx}{\linewidth}{
    @{}
    >{\raggedright\arraybackslash}p{3.55cm}
    >{\hsize=1.15\hsize\linewidth=\hsize\centering\arraybackslash}X
    >{\hsize=1.15\hsize\linewidth=\hsize\centering\arraybackslash}X
    >{\hsize=0.72\hsize\linewidth=\hsize\centering\arraybackslash}X
    >{\hsize=1.15\hsize\linewidth=\hsize\centering\arraybackslash}X
    >{\hsize=0.82\hsize\linewidth=\hsize\centering\arraybackslash}X
    >{\hsize=1.01\hsize\linewidth=\hsize\centering\arraybackslash}X
    @{}
}
\toprule
\textbf{Name} &
\shortstack{\scriptsize\textbf{Full-pipeline}\\\scriptsize\textbf{automation}} &
\shortstack{\scriptsize\textbf{Independent}\\\scriptsize\textbf{proving}\\\scriptsize\textbf{runtime}} &
\shortstack{\scriptsize\textbf{Built-in}\\\scriptsize\textbf{domain}\\\scriptsize\textbf{library}} &
\shortstack{\scriptsize\textbf{Proof-driven}\\\scriptsize\textbf{library}\\\scriptsize\textbf{evolution}} &
\shortstack{\scriptsize\textbf{Domain-}\\\scriptsize\textbf{specific}\\\scriptsize\textbf{modeling}} &
\shortstack{\scriptsize\textbf{Graphical}\\\scriptsize\textbf{task}\\\scriptsize\textbf{workspace}} \\
\midrule
\sysname{\textbf{Statopia Prover + Statlib}}{1} & $\checkmark$ & $\checkmark$ & $\checkmark$ & $\checkmark$ & $\checkmark$ & $\checkmark$ \\
\addlinespace[3pt]
\sysname{Stat-Lean}{2} & $\checkmark$ & -- & $\checkmark$ & $\checkmark$ & $\checkmark$ & -- \\
\sysname{KAPPA}{3} & $\checkmark$ & -- & -- & -- & $\checkmark$ & -- \\
\sysname{CausalSmith + Causalean}{4} & $\checkmark$ & -- & $\checkmark$ & $\checkmark$ & $\checkmark$ & -- \\
\sysname{ProofLoom + SOptLib}{5} & $\checkmark$ & -- & $\checkmark$ & $\checkmark$ & $\checkmark$ & -- \\
\addlinespace[3pt]
\sysname{AI4SLT / FoML / StatsMLlib}{6} & -- & -- & $\checkmark$ & -- & $\checkmark$ & -- \\
\sysname{stat-lib/statlib}{7} & -- & -- & $\checkmark$ & -- & $\checkmark$ & -- \\
\sysname{FormalSLT (v0.2.0)}{8} & -- & -- & $\checkmark$ & -- & $\checkmark$ & -- \\
\sysname{LML / LMLExplore}{9} & -- & -- & $\checkmark$ & -- & $\checkmark$ & -- \\
\sysname{FLTK}{10} & -- & -- & $\checkmark$ & $\checkmark$ & $\checkmark$ & -- \\
\sysname{OptLib / OptProver}{11} & -- & $\checkmark$ & $\checkmark$ & -- & $\checkmark$ & -- \\
\addlinespace[3pt]
\sysname{AutoformBot / ATLAS}{12} & $\checkmark$ & $\checkmark$ & $\checkmark$ & -- & $\checkmark$ & -- \\
\sysname{M2F / Quokka}{13} & $\checkmark$ & -- & -- & -- & -- & -- \\
\sysname{LeanFlow}{14} & $\checkmark$ & $\checkmark$ & -- & -- & -- & $\checkmark$ \\
\sysname{LeanMarathon}{15} & $\checkmark$ & -- & -- & -- & -- & -- \\
\sysname{Rethlas--Archon / Horizon}{16} & $\checkmark$ & -- & -- & $\checkmark$ & -- & $\checkmark$ \\
\sysname{OpenGauss}{17} & $\checkmark$ & -- & -- & -- & -- & -- \\
\sysname{OpenProver (API)}{18} & -- & $\checkmark$ & -- & -- & -- & -- \\
\sysname{Numina-Lean-Agent / Fuse}{19} & $\checkmark$ & -- & -- & -- & -- & $\checkmark$ \\
\sysname{PaperFormalizer (paper)}{20} & $\checkmark$ & -- & -- & $\checkmark$ & $\checkmark$ & -- \\
\sysname{DreamProver}{21} & -- & $\checkmark$ & -- & $\checkmark$ & -- & -- \\
\bottomrule
\end{tabularx}
\vspace{3pt}
\begin{minipage}{\linewidth}
\scriptsize
\textit{Sources (by row).}
\textsuperscript{1}\citep{statopia2026prover,statopia2026statlib};
\textsuperscript{2}\citep{wei2026hypothesis,statlean2026};
\textsuperscript{3}\citep{she2026kappa};
\textsuperscript{4}\citep{tan2026causalsmith};
\textsuperscript{5}\citep{wang2026proofloom};
\textsuperscript{6}\citep{zhang2026aislt,statsmllib2026,sonoda2026rademacher};
\textsuperscript{7}\citep{statlibcommunity2026};
\textsuperscript{8}\citep{formalslt2026};
\textsuperscript{9}\citep{degenne2026lml,lmlexplore2026};
\textsuperscript{10}\citep{gupta2026fltk};
\textsuperscript{11}\citep{li2024firstorder,li2026optprover};
\textsuperscript{12}\citep{rammal2026formalizing,autoformbot2026paperimplementation,autoformbot2026implementation,atlas2026v2};
\textsuperscript{13}\citep{wang2026m2f,m2f2026implementation};
\textsuperscript{14}\citep{milikic2026leanflow,leanflow2026implementation};
\textsuperscript{15}\citep{zhang2026leanmarathon};
\textsuperscript{16}\citep{ju2026automated,frenzymath2026archon,frenzymath2026horizon};
\textsuperscript{17}\citep{mathinc2026opengauss};
\textsuperscript{18}\citep{kripner2026openprover,openprover2026implementation};
\textsuperscript{19}\citep{liu2026numina,asher2026fuse};
\textsuperscript{20}\citep{tan2026paperformalizer};
\textsuperscript{21}\citep{zhang2026dreamprover}.
\par\smallskip
\textit{Reading the table.}
$\checkmark$ denotes documented support in at least one cited family member; -- means that none of the cited members establishes the capability under the stated criterion, including capabilities outside their scope. Family rows combine versions, implementations, companion projects and documented library lineages; shared authorship or research topic alone does not define a family.
A family's own standalone or companion domain library satisfies the third column.
An author-grown toolkit or a generated lemma collection can satisfy library evolution without constituting a built-in domain library.
AutoformBot and PaperFormalizer provide graphical monitoring and interaction, but their cited workflows start tasks outside that interface; their final-column marks do not mean that they lack a GUI.
\end{minipage}
\end{table}

\paragraph{General-purpose systems.}
These systems pursue several complementary directions. AutoformBot coordinates large-scale textbook formalization, producing the ATLAS collection \citep{rammal2026formalizing}, while M2F separates source translation into statement skeletons from subsequent proof completion \citep{wang2026m2f}. LeanFlow organizes source preparation, statement review and proof construction into a persistent workflow \citep{milikic2026leanflow}, LeanMarathon uses evolving blueprints and dependency graphs for long-horizon proof development \citep{zhang2026leanmarathon}, and Rethlas--Archon couples informal reasoning with Lean formalization and uses Matlas for mathematical literature retrieval \citep{ju2026matlas,ju2026automated}. Numina-Lean-Agent equips a general coding agent with Lean interaction and mathematical reasoning tools \citep{liu2026numina}; OpenGauss provides project-level orchestration \citep{mathinc2026opengauss}; and OpenProver combines parallel proof search, Lean verification and an interactive terminal interface, with formal verification starting from supplied Lean statements \citep{kripner2026openprover}.

More recent systems also make persistent human interaction and library feedback explicit. Archon Horizon coordinates long-running formalization through a graphical workspace and supports reviewed postprocessing of completed proofs into destination-library contributions \citep{frenzymath2026horizon}. Numina Fuse combines source papers, blueprints, dependency graphs, Lean code and agent interaction in a graphical workspace \citep{asher2026fuse}. PaperFormalizer starts from a paper's LaTeX source, involves its author in resolving ambiguities and substantive repairs, and promotes reusable results into the author's personal toolkit \citep{tan2026paperformalizer}. Its treatment-effect case study also examines statistical assumptions, including overlap and tail-moment conditions. Its selectable domain libraries and personal toolkit are distinct from a built-in project domain library. A complementary line studies reusable lemma accumulation: LEGO-Prover grows a verified skill library in Isabelle \citep{wang2024lego}, and DreamProver refines transferable Lean lemma libraries through alternating proving and consolidation stages \citep{zhang2026dreamprover}.

\paragraph{Statistics and machine learning.}
Closer to our setting, existing work develops specialized workflows and reusable formal infrastructure. Stat-Lean combines hypothesis-disciplined statistical formalization with a shared library and promotion of reusable intermediate lemmas \citep{wei2026hypothesis,statlean2026}. KAPPA emphasizes faithful checking of statistical paper proofs through statement review, explicit treatment of unproved inputs, and separate proof construction and audit \citep{she2026kappa}. CausalSmith builds an automated research pipeline on its Causalean library and promotes reusable lemmas proved during its runs back into that library \citep{tan2026causalsmith}. In stochastic optimization, ProofLoom constructs algorithm models and supporting theory from source proofs; its SOptLib accumulates verified results and modeling experience through extraction, generalization and later reuse \citep{wang2026proofloom}.

On the library side, AI4SLT develops empirical-process and concentration infrastructure \citep{zhang2026aislt}, while FoML formalizes generalization bounds through Rademacher complexity and entropy methods \citep{sonoda2026rademacher}. StatsMLlib brings together probability, statistics and learning-theoretic infrastructure and documents its integration of material from the former SLT and FoML trees \citep{statsmllib2026}. We therefore group AI4SLT, FoML and StatsMLlib as a library lineage, while preserving their separate source attributions. FormalSLT independently develops statistical learning and finite-sample inference \citep{sneiderman2026agents,formalslt2026}. The Formal Learning Theory Kernel organizes learning theory into reusable mathematical layers and documents proof-driven changes to definitions and shared interfaces \citep{gupta2026fltk}. Lean Machine Learning provides shared representations for machine learning theory \citep{degenne2026lml}; its AI-authored companion LMLExplore follows human-written roadmaps, with maintainers selecting material for incorporation into LML \citep{lmlexplore2026}. The independent \texttt{stat-lib/statlib} project emphasizes vocabulary and API design for mathematical statistics \citep{statlibcommunity2026}.\footnote{Despite the similar name, this project is unrelated to Statlib, the Statopia library developed in this work.} OptLib formalizes optimization assumptions and convergence theory \citep{li2024firstorder}, while OptProver builds its optimization experiments on OptLib and adapts proof models to already formalized optimization goals \citep{li2026optprover}. Their shared row attributes domain infrastructure to OptLib and the proving runtime to OptProver.

\paragraph{Comparison criteria.}
Full-pipeline automation covers source formalization and proof construction, with user review or guidance permitted. Runtime independence means that the proving loop executes without an external coding-agent runtime; calls to model APIs are permitted. The built-in domain-library column requires a project's own maintained collection of domain definitions and theorem APIs, either integrated into the system or supplied as its companion library. Connecting an external library alone does not meet this criterion. Proof-driven library evolution requires proof or reuse evidence to guide the extraction, generalization or revision of shared formal results and interfaces; it may involve human review. Routine theorem contributions, project growth, and model or prompt updates alone do not establish this capability. Domain-specific modeling concerns statistical, learning-theoretic or optimization representations and assumption handling, rather than specialized model training or generic statement checking alone. The workspace criterion requires graphical task submission, monitoring and execution control; terminal interfaces and result viewers alone do not qualify.

\paragraph{Versions and configurations.}
The family grouping was updated on 8 November 2026. Capability evidence uses the paper versions and software snapshots checked for the 7 November comparison; the StatsMLlib lineage was additionally checked on 8 November. Versions and sources are specified in the references. The AutoformBot / ATLAS family covers the published API-based pipeline (\texttt{v0}), the later coding-agent plugin, and the companion ATLAS library \citep{autoformbot2026paperimplementation,autoformbot2026implementation,atlas2026v2}. Its independent-runtime mark comes from the original pipeline; its domain-library and modeling marks come from the cited ATLAS v2 development snapshot, which includes maintained statistical definitions and theorem APIs. LeanFlow combines the paper's direct-API configuration (runtime independence) with v0.3.0 (graphical workspace); these marks need not hold together in either implementation. The Archon family combines Rethlas--Archon v0.3.3 with Archon Horizon v0.2.0-alpha.2 at the cited revision: both document library feedback, and Horizon supplies the graphical workspace. The Numina family combines Numina-Lean-Agent with the hosted Fuse system described in its paper; the workspace mark comes from Fuse. Fuse Desktop is mentioned in that paper but is not separately evaluated here.

The M2F / Quokka row includes the paper and its named online engineering implementation \citep{m2f2026implementation}; its marks rely on the verified paper workflow, and the online-service description alone does not establish the graphical task controls required by the final column. LML / LMLExplore combines the core library and its AI-authored companion; human selection of contributions alone does not establish proof-driven library evolution. OptLib / OptProver combines the optimization library with the related prover evaluated on it; its independent-runtime mark follows OptProver's LeanDojo and best-first-search setup, while the domain marks refer to OptLib. PaperFormalizer denotes arXiv v2 and its author-toolkit workflow. OpenProver's runtime mark refers to its API configuration.

These systems already provide several of the capabilities relevant to Statopia. Our design brings a native proving runtime, statistical domain infrastructure, reviewed feedback between proving and library development, and a graphical task workspace into one system. Owning the proving loop gives direct control over context management, tool scheduling, Lean verification and recovery, including their interaction with retrieval and theorem reuse. This architectural distinction motivates joint design of the prover and Statlib; the capability comparison alone does not establish a performance advantage over systems built on general-purpose coding-agent runtimes.

\section{Statopia Prover}
\label{sec:overview}

This section first gives a brief overview of the architecture of Statopia Prover and of how it is used, and then describes the main components of its workflow: preparation and formalization, parallel proof development with proof DAGs, and task groups.

\subsection{Architecture and Use}
\label{sec:architecture}
\label{sec:using-statopia}

\begin{figure}[!t]
\centering
\resizebox{0.98\linewidth}{!}{%
\begin{tikzpicture}[
  >={Latex[length=2mm]},
  every node/.style={align=center,font=\small,
    execute at begin node={\hyphenpenalty=10000\exhyphenpenalty=10000}},
  comp/.style={draw,rounded corners=1.5mm,thick,minimum height=17mm,text width=33mm,inner sep=3pt},
  base/.style={draw,rounded corners=1.5mm,thick,minimum height=13mm,text width=46mm,inner sep=3pt},
  flow/.style={->,thick,draw=gray!70!black},
  dflow/.style={->,thick,dashed,draw=gray!70!black},
  lbl/.style={font=\footnotesize,fill=white,inner sep=1.5pt},
  band/.style={rounded corners=3mm,very thick,inner sep=8pt},
  bandtitle/.style={anchor=south west,font=\bfseries\small,fill=white,inner sep=1.5pt}
]
\node[comp,draw=gray!65!black,fill=gray!8] (src) at (0,9.0)
  {\textbf{Sources}\\papers\\theorem descriptions\\Lean developments};
\node[comp,draw=blue!65!black,fill=blue!8,text width=60mm] (ws) at (6.0,9.0)
  {\textbf{Workspace}\\task groups \(\supset\) tasks \(\supset\) proof DAGs\\review of formal targets \(\cdot\) progress};
\node[comp,draw=gray!65!black,fill=gray!8] (art) at (12.0,9.0)
  {\textbf{Verified artifacts}\\Lean package\\verification status\\citation inputs};
\node[comp,draw=cyan!60!black,fill=cyan!9] (prep) at (0,5.6)
  {\textbf{Preparation}\\organize sources\\formalize target\\plan the proof};
\node[comp,draw=violet!65!black,fill=violet!8] (coord) at (4.0,5.6)
  {\textbf{Task groups}\\main task and sub-tasks\\shared results};
\node[comp,draw=blue!65!black,fill=blue!9] (dag) at (8.0,5.6)
  {\textbf{Proof DAGs}\\lemmas as nodes\\parallel development};
\node[comp,draw=red!65!black,fill=red!7] (gate) at (12.0,5.6)
  {\textbf{Verification}\\locked statement\\Lean build, no \texttt{sorry}\\integrity checks};
\draw[flow] (prep) -- (coord);
\draw[flow] (coord) -- (dag);
\draw[flow] (dag) -- (gate);
\node[base,draw=green!50!black,fill=green!12] (statlib) at (0.65,2.4)
  {\textbf{Statlib}\\checked declarations};
\node[base,draw=orange!75!black,fill=orange!10] (lean) at (6.0,2.4)
  {\textbf{Lean~4 and Mathlib}\\on a local or cloud host};
\node[base,draw=purple!60!black,fill=purple!7] (llm) at (11.35,2.4)
  {\textbf{Language model APIs}\\replaceable backends\\user's own key};
\begin{scope}[on background layer]
  \node[band,draw=blue!60!black,fill=blue!2,fit=(src)(ws)(art)] (users) {};
  \node[band,draw=red!65!black,fill=red!1,fit=(prep)(coord)(dag)(gate)] (engine) {};
\end{scope}
\node[bandtitle,text=blue!60!black] at ([xshift=2mm]users.north west) {Users and workspaces};
\node[bandtitle,text=red!65!black] at ([xshift=2mm]engine.north west) {Statopia Prover};
\draw[flow] (src) -- node[lbl,above] {submit} (ws);
\draw[dflow] (art) -- node[lbl,above] {reuse} (ws);
\draw[flow] (ws.south |- users.south) -- node[lbl,right] {task groups} (ws.south |- engine.north);
\draw[flow] (gate.north |- engine.north) -- node[lbl,right] {verified results} (art.south |- users.south);
\draw[flow] ([xshift=-8mm]statlib.north) -- node[lbl,left] {retrieve} ([xshift=-8mm]statlib.north |- engine.south);
\draw[dflow] ([xshift=8mm]statlib.north |- engine.south) -- node[lbl,right] {reviewed promotion} ([xshift=8mm]statlib.north);
\draw[flow,<->] (lean.north) -- node[lbl,right] {compile \(\cdot\) check} (lean.north |- engine.south);
\draw[flow,<->] (llm.north) -- node[lbl,right] {proposals} (llm.north |- engine.south);
\end{tikzpicture}}
\caption{System overview of Statopia. Users submit sources, with or without proofs, to a workspace, which organizes formalization work into task groups, tasks and proof DAGs. Statopia Prover prepares each task, develops its proof and accepts results only after the final verification checks. Statlib supplies checked declarations for reuse and receives verified results only through reviewed promotion (dashed). Lean runs on a local or cloud host, and language models are accessed through replaceable APIs.}
\label{fig:system-overview}
\end{figure}
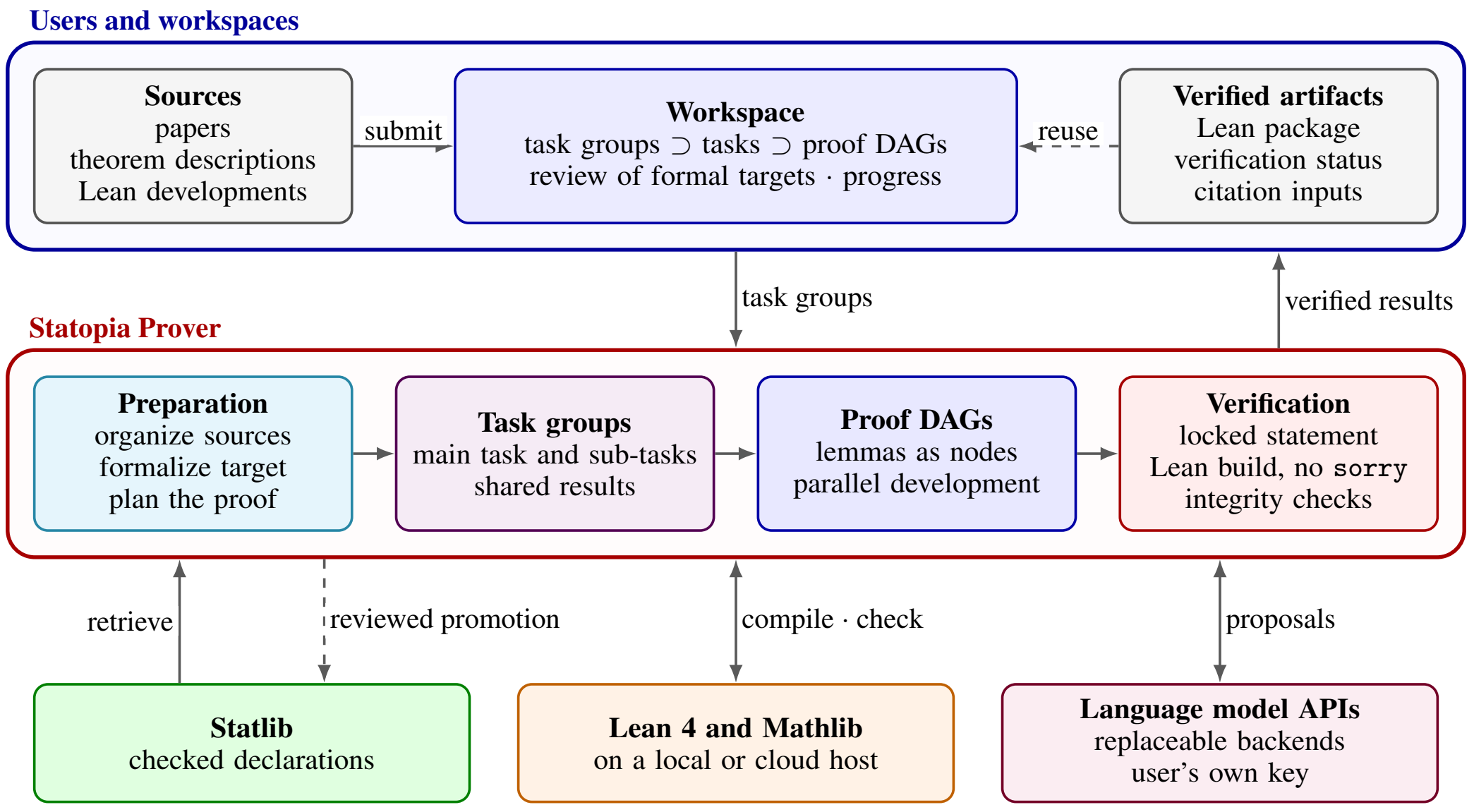

Statopia Prover is a desktop application for macOS and Windows, available at \url{https://prover.statopia.ai/download}. Figure~\ref{fig:system-overview} shows its main parts and how they interact. Users submit sources to workspaces, where formalization work is organized into task groups, tasks and proof DAGs (Section~\ref{sec:task-groups}). The prover runs inside the desktop application, in its own execution harness rather than inside a general-purpose coding agent, and it manages the working context of its agents itself. Language models are replaceable backends: they propose formalizations, proof plans and Lean edits, while the prover decides which proposals take effect and Lean checks every proof. Lean runs with Mathlib and Statlib on the user's machine or on a cloud host, and cloud services keep account, workspace and result records. Statlib is built for use by the prover and is itself developed with it. It supplies checked declarations for reuse and receives new verified results only through a separate reviewed promotion process (Section~\ref{sec:statlib}).

After signing in, users add an API key for their chosen language model provider. The key is stored encrypted on their own machine, and during local runs model calls are sent from that machine rather than through Statopia's servers. A typical session has five steps.

\begin{description}
    \item[Submit.] Create a workspace and add a paper or other source document, a theorem description or an existing Lean development. A source need not contain a proof: a theorem description may state only the result, and the prover then develops the proof itself (Section~\ref{sec:statopia-prover}). A workspace can hold several related theorems, and verified results from earlier tasks in the same workspace can support later ones.
    \item[Review the formal target.] Statopia proposes a Lean statement together with the assumptions it reconstructed from the source. The user approves it, clarifies an ambiguity, or revises the source and re-imports it.
    \item[Prove locally or in the cloud.] Once the local Lean environment passes its checks, proofs run on the user's machine; while it is still being installed, a cloud route can take over when capacity allows. Even when proofs run locally, the state of each task is synchronized to Statopia's servers; this synchronization and model access always require an internet connection.
    \item[Monitor and resume.] The workspace shows the progress of each task and its unresolved obligations. Developments can be paused and later resumed from their checkpoints. In the case of certain unexpected interruptions, a task can also resume from its most recent checkpoint rather than start over.
    \item[Inspect the result.] A result is marked ``Proved'' only if the proof of its locked statement builds in Lean without \texttt{sorry}, passes the integrity checks, relies only on explicitly declared assumptions and uses no prohibited axioms. The result record keeps the Lean package, its build outcome, any unresolved obligations and every retained citation input, so users can see exactly what each claim rests on. Optional narration and a PDF report can be generated and exported for a task or a task group.
\end{description}

The online Statlib catalog lets users browse the library by topic; as of 1 November 2026 it lists 1,807 theorems and 1,035 supporting lemmas in 470 Lean files. Statopia serves researchers in statistics and machine learning who want to verify results from familiar sources or to obtain formal proofs of their own claims, even without a proof at hand; Lean formalizers who need statistical infrastructure; and library developers who can use evidence from proof development to improve Statlib.

\subsection{Preparation and Formalization}
\label{sec:statopia-prover}
\label{sec:material-preparation}

A theorem in statistics or machine learning often depends on a model, an algorithm, an estimator and several layers of assumptions whose definitions are spread across the paper, its appendices and cited works. Its formal statement may also require properties left implicit in the exposition, such as measurability, integrability or the existence of the quantities being optimized. Before any proof is attempted, Statopia Prover therefore organizes the relevant statements, definitions, assumptions and any proof passages from the sources and keeps track of where each one comes from.

Specialized agents then formalize the target theorem and its supporting definitions in Lean, compile the resulting statement and review it for its correspondence with the source; the user reviews the formal target as well (Section~\ref{sec:using-statopia}). Once the target theorem is formalized, its Lean statement is locked: proof-writing agents cannot change it, so a proof cannot succeed by quietly weakening the theorem, and any correction to the statement goes through a separate review.

Starting from the organized material, the prover develops a natural-language argument to guide the Lean proof. If the sources contain a proof, the prover fills in the steps that the proof omits and may propose an alternative route; if they state the result without proof, the prover devises the argument itself. Either way, the argument makes intermediate claims and the use of results from the literature explicit. A route assessment then examines how this argument can be realized in Lean, using library declarations whose exact signatures have been checked, and refines it into a proof guide. If the assessment reveals a mismatch with the source or a gap in the argument, the task returns to the corresponding step; such corrections preserve the intended theorem rather than add assumptions or weaken the conclusion. Preparation ends with a formal target, a proof guide and an initial proof DAG of intermediate lemmas.

\paragraph{Citation inputs.}\label{sec:engine-overview} Keeping cited results as inputs by default, rather than proving them, is a deliberate design choice. Statistical papers often build on results from the literature, such as a classical inequality or a theorem from an earlier paper, whose proofs can be long and lie outside the paper's own contribution. Proving all of them would make each development much larger and shift effort away from the paper's own arguments. When such a result is not available in Mathlib or Statlib, it is instead stated in Lean as a citation input, recorded with its source and listed in the result record. This keeps each development at a manageable size, mirrors how a paper itself relies on earlier work and makes explicit which cited results a verified theorem depends on. The default also leaves the choice to the user: those who prefer a self-contained development can have Statopia Prover prove all citation inputs as well, so that the final result no longer depends on them. Citation inputs enter the proof DAG as external nodes (Section~\ref{sec:dynamic-dag}).

\subsection{Proving in Parallel with Proof DAGs}
\label{sec:dynamic-dag}

Relatively simple targets can be handled by a single-file proof procedure, and larger targets are developed through a dynamic proof DAG whose nodes are the target theorem and its intermediate lemmas and whose edges record which results each proof uses. A node can also be external to the task. It then provides a statement that the task uses without proving it: a theorem from Statlib, which is already verified; the result of another task, which counts only once its own proof is accepted; or a citation input. Each external node records where its statement comes from, so that relying on it is not mistaken for having proved it in the current task.

The proof DAG is developed in parallel. Separate agents work on different nodes at the same time, each concentrating on one bounded obligation with the material relevant to its current step, so a large proof advances on many fronts at once instead of one lemma after another. Independent parts of the development proceed without waiting for one another, which shortens the elapsed time of long developments, and records of earlier attempts let a new attempt build on useful partial work instead of repeating a failed approach. Section~\ref{sec:individual-evaluation} reports results of this dynamic DAG procedure on 42 benchmark problems.

The graph also changes as the proof develops: a substantive intermediate result can become a new helper lemma, and a lemma that turns out to be unsound or poorly suited to the proof can be revised or withdrawn. These changes are reviewed before they take effect, only the work that depends on a changed result is redone, and new helpers remain proof obligations of the task rather than assumptions.


\subsection{Task Groups and Organization}
\label{sec:task-groups}
\label{sec:work-organization}

Statopia organizes formalization work at four nested levels. A \emph{workspace} is a project: it holds the user's sources, instructions and references, the theorems to be formalized, and the verified results that later work can reuse. Within a workspace, each target theorem is developed as a \emph{task group}, consisting of a main task for the theorem and sub-tasks for supporting results that the source states separately, such as the lemmas of a paper. A \emph{task} is a complete proof problem with its own formal target, proof guide and recorded progress, and within each task the remaining proof obligations form a dynamic \emph{proof DAG} (Section~\ref{sec:dynamic-dag}). This organization mirrors how statistical papers are written and keeps each proof problem at a manageable size.


Task groups let users work with a paper the way it is written. Statopia identifies the lemmas and other supporting results that the paper states separately and, where appropriate, sets them up as sub-tasks that follow the same workflow as the main task. A lemma used by several results is proved once and shared, and verified results from earlier tasks in the workspace can support later ones. In the workspace, users can follow the progress of each task and its unresolved obligations and inspect its result, and narration and PDF reports can be exported for a single task or for a whole task group. A paper-scale development can therefore be followed and checked lemma by lemma, while the main theorem is marked ``Proved'' only when the tasks it relies on are verified as well. Section~\ref{sec:paper-evaluation} reports paper developments organized in this way, with up to 16 tasks in a group.

If the statement of a shared lemma needs correction, the correction is reviewed against the source and passed on to the tasks that use the lemma, and verified work that remains valid is kept. Because a paper-scale development can run for hours, Statopia saves the progress of a task group as it goes: users can pause the group and resume it later from where it stopped, and a running group can also move between their machine and the cloud.

\section{Statlib: Reusable Statistical Infrastructure}
\label{sec:statlib}
Statlib is the reusable statistical infrastructure of the Statopia ecosystem. It is built as a domain-specific API on top of Mathlib, not as a collection of isolated formalizations. It gives statistical concepts stable Lean representations, packages recurring proof steps behind explicit interfaces and organizes related theorems into families that can be reused directly, both in manual formalization and in automated proof construction. The whole library is developed with Statopia Prover, and its modules are reviewed before they are added (Section~\ref{sec:library-evolution}).

This organization matters especially for automated proving. Statistical arguments often pass through several mathematical layers, for example from measurability and moment assumptions to a concentration bound, and then from that bound to a risk or consistency result. If different developments encode the same statistical concept in incompatible ways, the prover ends up rebuilding these connections again and again. Likewise, a formally correct result is of limited use if it is hard to find, import or combine with other results. Statlib therefore emphasizes consistent representations of statistical concepts, reusable results at several levels of granularity and a public interface that the prover can search and check efficiently.

\subsection{Layered Statistical Library Design}

Mathlib provides the general formal foundations: measure theory, probability, analysis, topology, algebra and linear algebra \citep{mathlib2020}. On top of these, Statlib adds the domain-specific infrastructure needed to state and reuse results in statistics and machine learning. Its current top-level areas include statistical foundations, high-dimensional statistics, nonparametric statistics, hypothesis testing, random matrix theory and causal inference \citep{statopia2026statlib}. These areas may differ in maturity and coverage, but they share the same vocabulary, interface conventions and verification infrastructure.

\begin{figure}[!htbp]
\centering
\resizebox{0.98\linewidth}{!}{%
\begin{tikzpicture}[
  y=0.72cm,
  >={Latex[length=2.2mm]},
  every node/.style={align=center, font=\scriptsize},
  box/.style={
    draw,
    rounded corners=1.5mm,
    thick,
    minimum height=8mm,
    inner sep=3pt
  },
  widelayer/.style={box, text width=6.5cm},
  halflayer/.style={box, text width=3.2cm, minimum height=10mm},
  accessnode/.style={box, text width=3.7cm, minimum height=10mm},
  statlibgroup/.style={
    draw=green!55!black,
    fill=green!3,
    rounded corners=3mm,
    very thick,
    inner sep=7pt
  },
  outer/.style={draw=red!65!black,rounded corners=3.5mm,
    very thick,inner sep=8pt}
]

\node[widelayer, draw=blue!65!black, fill=blue!10] (vocab) at (0,0) {
  \textbf{Vocabulary and interfaces}\\
  {\scriptsize objects $\cdot$ predicates $\cdot$ assumptions}
};

\node[halflayer, draw=cyan!60!black, fill=cyan!12] (atomic) at (-2.0,1.95) {
  \textbf{Operational lemmas}\\
  {\scriptsize closure $\cdot$ side conditions\\normalization}
};

\node[halflayer, draw=orange!75!black, fill=orange!13] (bridges) at (2.0,1.95) {
  \textbf{Bridge lemmas}\\
  {\scriptsize conversions $\cdot$ equivalences\\interface links}
};

\node[widelayer, draw=green!55!black, fill=green!12] (families) at (0,4.1) {
  \textbf{Reusable theorem families}\\
  {\scriptsize concentration $\cdot$ convergence $\cdot$ risk $\cdot$ matrix and spectral bounds}
};

\node[widelayer, draw=violet!65!black, fill=violet!10] (topics) at (0,6.0) {
  \textbf{Topic and application modules}\\
  {\scriptsize coherent domain entry points and theorem chains}
};

\node[box, draw=gray!70!black, fill=gray!10, text width=5.7cm]
  (mathlib) at (0,-2.3) {
  \textbf{Mathlib foundations}\\
  {\scriptsize measure theory $\cdot$ probability\\analysis $\cdot$ algebra}
};

\node[accessnode, draw=orange!75!black, fill=orange!9] (search) at (6.8,3.0) {
  \textbf{Prover-facing search surface}\\
  {\scriptsize names $\cdot$ signatures\\
  locations $\cdot$ curated aliases}
};

\node[accessnode, draw=red!60!black, fill=red!8] (prover) at (6.8,6.0) {
  \textbf{Statopia Prover}\\
  {\scriptsize retrieve $\rightarrow$ check $\rightarrow$ compose}
};

\begin{scope}[on background layer]
  \node[statlibgroup, fit=(vocab)(atomic)(bridges)(families)(topics)]
    (statlib) {};
\end{scope}

\node[anchor=south west, font=\bfseries\scriptsize,text=green!50!black,
  fill=white,inner sep=1.5pt]
  (statliblabel) at (statlib.north west) {Statlib --- reusable statistical API};

\draw[->, thick] (mathlib.north) -- (vocab.south);
\draw[->, thick, blue!65!black] (atomic.south |- vocab.north) -- (atomic.south);
\draw[->, thick, blue!65!black] (bridges.south |- vocab.north) -- (bridges.south);
\draw[->, thick, cyan!60!black] (atomic.north) -- (atomic.north |- families.south);
\draw[->, thick, orange!75!black] (bridges.north) -- (bridges.north |- families.south);
\draw[->, thick, green!55!black] (families.north) -- (topics.south);

\draw[->, thick, dashed, orange!75!black]
  (statlib.east |- search.west) -- (search.west);
\draw[->, thick, red!60!black] (search.north) -- (prover.south);

\begin{scope}[on background layer]
  \node[outer,fit=(statlib)(statliblabel)(mathlib)(search)(prover)] (architecture) {};
\end{scope}
\node[anchor=south west,font=\bfseries\scriptsize,text=red!65!black,
  fill=white,inner sep=1.5pt]
  at ([xshift=2mm]architecture.north west) {Statlib: layered statistical infrastructure};

\end{tikzpicture}}
\caption{Layered architecture of Statlib and its prover-facing search surface. Mathlib provides the general mathematical foundations. The statistical vocabulary supports operational and bridge lemmas, which combine into reusable theorem families and domain modules. Declaration data and curated aliases make these components retrievable by Statopia Prover. Retrieved declarations are checked before they are used in proof planning and construction, and final proof acceptance still depends on Lean elaboration, the build and the integrity checks.}
\label{fig:statlib-layered-architecture}
\end{figure}
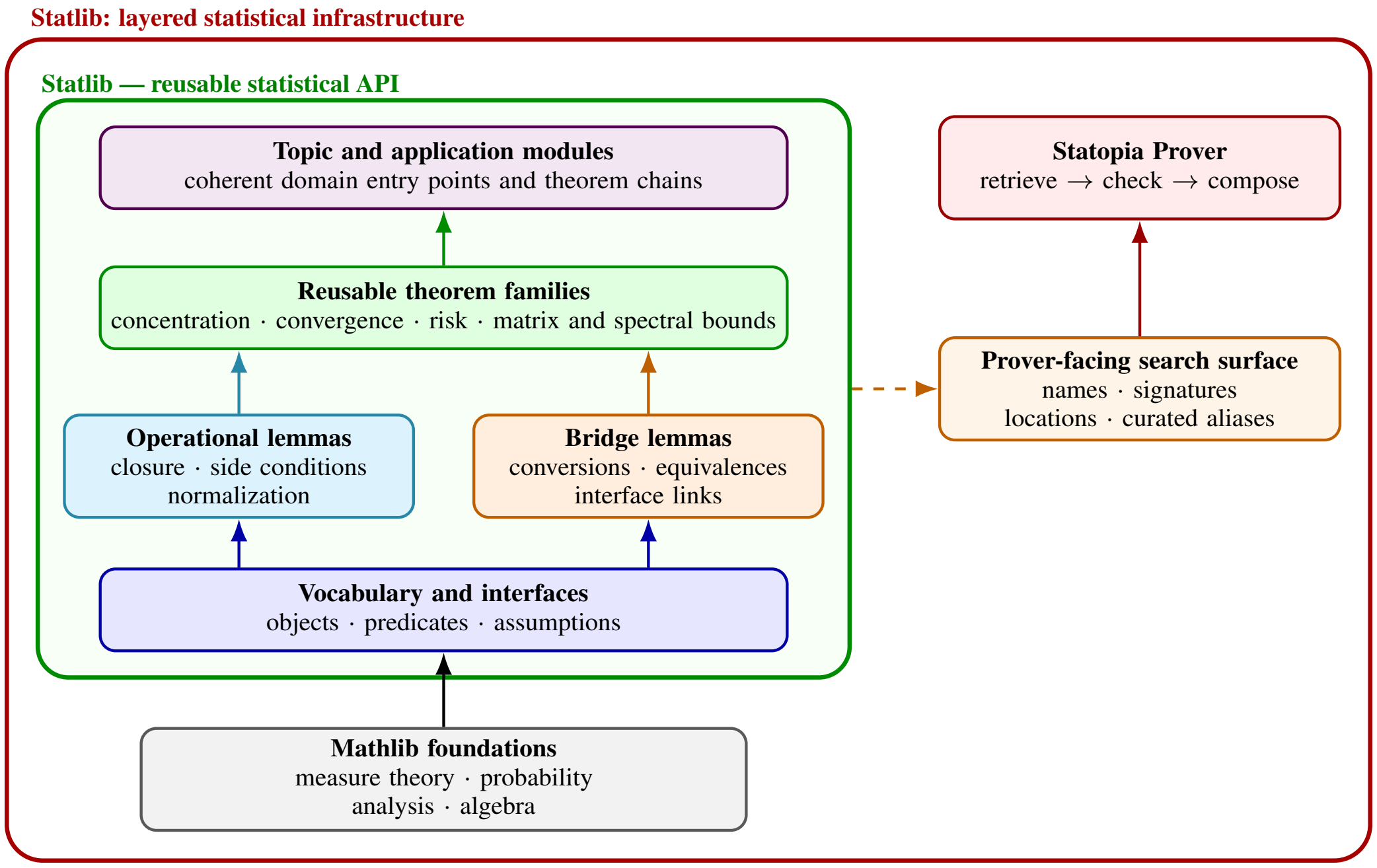

Figure~\ref{fig:statlib-layered-architecture} shows how these components are organized. Each layer is defined by the role its declarations play in formalization and proof reuse. A stable statistical vocabulary forms the common representation layer; operational and bridge lemmas connect these representations to recurring proof steps; and larger theorem families are assembled into coherent topic and application modules. Statopia Prover reaches this whole structure through a prover-facing search surface, shown on the right of the figure, which it uses to retrieve and reuse declarations.

The vocabulary and interface layer gives each statistical concept a stable public representation in Lean. For example, an informal i.i.d. assumption combines independence and identical distribution, and a concentration theorem may also require measurability, centering, integrability or a condition on the moment generating function. Expressing these requirements through shared predicates and structures makes the assumptions explicit and lets different theorem families reuse them. The same approach covers sparsity conditions, smoothness classes, risks, estimators, test procedures, covariance assumptions and causal identification criteria. With these interfaces, the formalizer and the prover can compare the theorem they are working on with the exact assumptions of a candidate result.

Operational and bridge lemmas connect this vocabulary to larger mathematical arguments. Operational lemmas handle recurring local steps such as closure properties, measurability, normalization and other side conditions. Bridge lemmas connect equivalent or closely related representations, so results stated through different interfaces can be combined without rebuilding the conversion each time. Above these layers, reusable theorem families package results such as concentration bounds, convergence statements, risk bounds and matrix or spectral inequalities. Topic and application modules then arrange these results into coherent chains of theorems for each domain. The aim of this structure is to let results be combined across statistical developments, rather than to store final theorems as isolated artifacts.

\subsection{Reusable Granularity and Module Organization}
\label{sec:statlib-modules}

Statlib organizes definitions and theorems by mathematical topic. They sit under domain namespaces such as \path{Statlib.StatFoundation}, \path{Statlib.HighDim} and \path{Statlib.Nonparametric}, with narrower submodules for coherent theorem families. Topic entry modules give convenient access to a whole area, while individual developments import as few modules as is practical. This keeps elaboration contexts manageable and makes the mathematical dependencies of each development explicit.

Within this structure, a proof DAG produced by the prover is usually organized into a coherent Lean source file. The file collects the main theorem and the supporting declarations used in its proof. It makes public the main theorem and the important intermediate results selected for reuse, and keeps private the auxiliary declarations needed only inside the proof. The definitions needed to state and use the public results remain available, either through the shared vocabulary or through the file's public interface. Related proof steps thus stay together, and later developments see a focused set of results.

Source files and public declarations therefore serve different purposes. A file organizes a coherent proof development and is imported as a module, whereas each of its public declarations can be retrieved, checked and applied on its own in later proofs. One file may expose several such results, each with its own assumptions and conclusion. Shared vocabulary modules complement these developments with stable definitions, structures, classes and predicates that are used across many theorem families. How a development is split into files follows its mathematical structure and shared dependencies, so proof DAGs and source files need not correspond one to one.

Reusable public declarations come in several sizes. Small operational lemmas handle recurring local conditions, bridge lemmas connect compatible representations, and larger theorems package substantial mathematical arguments such as concentration inequalities, approximation bounds or spectral perturbation results. Whether a declaration suits public reuse depends on its mathematical role and on how stable its interface is. A small fact may remove an obligation that recurs at the Lean level, while a larger result may supply the central step of a later statistical argument. Arranging these declarations into coherent chains of theorems supports the reuse of both local proof steps and substantial statistical results.

\subsection{Prover-Native Retrieval and Verified Reuse}

Statlib is organized so that the prover can find and reuse its public declarations directly during automated proving. Naming is part of this interface, because it lets a statistical concept or theorem family be recognized both from its mathematical terminology and from its Lean declaration. Vocabulary declarations name the objects and assumptions used to state later results. Bundled structures, classes and named object types generally use PascalCase, value-level operations may use lowerCamelCase, and predicates commonly use PascalCase names beginning with \texttt{Is} or \texttt{Has}, as in \path{IsSubGaussianVector} and \path{HasBoundedSpectralNorm}. These names are meant to describe stable domain concepts rather than individual source papers.

Public theorem and lemma declarations name reusable propositions. Their names usually follow the snake-case conventions of Lean and Mathlib, which show the main object, the form of the conclusion and the main assumptions. Components such as \texttt{\_eq}, \texttt{\_le} and \texttt{\_iff} indicate the type of conclusion, while \texttt{\_of\_} introduces a condition under which the result holds. For example, \path{subexp_mgf_le_of_bounded} names both the bound the theorem provides and the assumption used to obtain it. A proof file may expose its main theorem and selected intermediate results, each under its own declaration name, while their namespace records the mathematical topic. With this convention, a concept-oriented query can locate the relevant vocabulary, and a theorem-oriented query can combine object names with the desired conclusions or assumptions.

The search surface is generated from the Lean source rather than maintained as a separate description of the library's contents. The linked library is compiled into Lean artifacts, and the prover builds an index of public declaration names, kinds and source locations, together with auxiliary information on signatures and definitions. Freshness checks based on content hashes detect when the linked source, the compiled library and the derived search data have drifted apart. A successful refresh rebuilds the linked library and regenerates the search data from the same snapshot of the source. If refreshing is disabled or fails, the system reports that the search data are out of date. The index thus makes retrieval efficient, while Lean remains the authority on declaration types, imports, elaboration and proof validity.

Search accepts approximate queries, because the terminology of papers and textbooks does not always match Lean identifiers. The local search layer handles exact names, name fragments, namespaces and multi-word concept queries over the combined Mathlib and Statlib surface. It normalizes common variants in spelling and separators, and uses a curated table of statistical synonyms for terms and abbreviations such as ``sub-Gaussian'', ``moment generating function'' and ``restricted isometry''. Candidate declarations are ranked by weighted coverage of the concepts in the query, and edit-distance suggestions help recover from a misspelled identifier or a wrong namespace. These procedures widen retrieval without treating textual similarity as mathematical evidence.

Before retrieval results are used for route planning, they are checked mechanically. The search process sends complementary queries based on mathematical terminology, assumptions and the likely form of Lean identifiers. Duplicate names are removed, and the exact declarations of the remaining names are checked in batches in the active Lean environment. A selection step then picks the useful evidence from the checked signatures; both the selected evidence and a size-limited log of the checked declarations are kept for planning, auditing and resumption. The Route Agent therefore reasons from verified declaration signatures, not from guessed theorem names or search snippets.

With these checked signatures, the Route Agent can compare the assumptions of candidate results with the current target, identify missing bridge obligations and choose a proof route that the available Lean API supports. Once the route is fixed, the formalizers and writers of individual nodes receive the relevant evidence and carry out only the narrower name, scope and type checks that their local obligations require, instead of restarting broad library research inside the writing loop. This division of labor lets the whole task reuse the results of expensive retrieval, and keeps local proof editing from silently replacing the mathematical route chosen during planning.

Throughout this process, retrieval is kept separate from proof acceptance. A candidate returned by approximate search becomes usable as proof evidence only after its exact declaration has been checked. Even a checked signature contributes to the theorem being proved only when its application elaborates in the current context and the resulting development passes Lean verification. Likewise, a limited query that returns nothing does not show that a theorem is missing from Mathlib or Statlib; it records only that this particular search produced no selected evidence. An unresolved mathematical requirement therefore stays part of the proof problem, instead of being turned into a claim about what the library contains.

\subsection{Library Evolution}
\label{sec:library-evolution}

Proof development shows directly how Statlib is used in practice and where more library support may help. Frequently selected declarations, recurring vocabulary queries, repeated side conditions and common bridge obligations can reveal which interfaces work well and where the current library repeatedly causes difficulties during formalization and proving.

Such evidence is interpreted in the context of the proof in which it arises. A helper lemma that appears repeatedly may capture a reusable mathematical fact, but it may also depend on a representation specific to one theorem. A type mismatch may point to a missing bridge lemma, or simply to an unsuitable proof route. Similarly, a search may fail because of terminology, ranking or import visibility rather than because the desired result is missing. Usage records and logs of checked searches are therefore treated as evidence for library development, not as automatic signs for evolution of Statlib.

Once a proof is complete, its proof DAG can be organized into a Statlib source file through the existing post-proof promotion path. This step keeps the proof and its supporting declarations, makes public the main theorem and the intermediate results worth reusing, and keeps the other auxiliary declarations private. Whether the file then enters the shared library is decided by the library's contribution and release rules. 

Evidence accumulated over many proofs can thus guide later Statlib development, by identifying useful additions, recurring needs for bridge lemmas and opportunities to improve existing interfaces. In this way, the library can evolve in response to concrete needs observed during formalization and proving, while staying separate from the verification of individual proofs. 



\section{Evaluation and Case Studies}
\label{sec:evaluation}

We evaluate Statopia in two settings: individual theorem tasks and developments of statistical papers. The first setting combines a statistics benchmark, in which no problem comes with a proof, with six guided IMO 2026 tasks. The second examines how the task groups for whole papers organize supporting results, maintain interfaces and produce integrated Lean artifacts. We also describe the case of a neural network approximation development, produced with Statopia Prover and organized into reusable Statlib modules. All Statopia runs reported here use \texttt{deepseek-v4-flash}. Costs are estimated from the recorded token usage at a single off-peak tariff: \$0.15 per million uncached input tokens, \$0.003 per million cached input tokens and \$0.60 per million output tokens.\footnote{DeepSeek pricing: \url{https://api-docs.deepseek.com/quick_start/pricing/}, accessed October 30, 2026.} They are therefore normalized estimates, not historical invoices, and usage records that could not be resolved are excluded. The current evaluation host has an Intel Xeon Platinum 8368 CPU at 2.40\,GHz. The results and their Lean source code are available in the public repository \url{https://github.com/statopia/established-results}.

\subsection{Individual Theorem Tasks}
\label{sec:individual-evaluation}

\paragraph{Statistics benchmark.}
Our benchmark draws on area~62 (Statistics) of the Mathematics Subject Classification and covers twelve branches: statistical foundations, decision theory, sampling and missing data, distribution theory, parametric and asymptotic inference, nonparametric methods, multivariate analysis, regression, experimental design, time series, survival analysis, and functional and directional data. The examples range from unbiasedness and finite-population variance calculations to central limit theorems, kernel estimation, principal component analysis and regularized regression. The proving track used here starts from fixed Lean statements, and no proof is supplied, either in natural language or in Lean. It therefore evaluates formal proof discovery: for each problem, the prover must devise the mathematical argument and construct the Lean proof itself. It does not evaluate the extraction of statements from natural-language documents.

The statistics benchmark consists of 83 distinct problems. The 41 relatively simple ones are handled by the single-file proof procedure, and the other 42 by the dynamic DAG procedure. All 83 problems were verified on their first end-to-end proving run, a first-run success rate of 100\% (83/83). Table~\ref{tab:statistics-results} summarizes these first-run results, one per problem, and the accepted Lean proofs are archived for inspection.

\begin{table}[!htbp]
\centering
\caption{Statistics benchmark results.}
\label{tab:statistics-results}
\small
\setlength{\tabcolsep}{4pt}
\begin{tabular}{@{}lrrrrr@{}}
\toprule
Proof procedure & Problems & Lean lines & Mean min. & Mean USD & USD/line \\
\midrule
Single file & 41 & 5,192 & 32.81 & 0.22 & 0.00174 \\
Dynamic DAG & 42 & 32,510 & 89.80 & 0.90 & 0.00117 \\
Combined & 83 & 37,702 & 61.65 & 0.57 & 0.00125 \\
\bottomrule
\end{tabular}
\end{table}

The 42 runs with the dynamic DAG procedure contain 266 verified DAG nodes: 42 main nodes and 224 auxiliary nodes that the prover introduced as intermediate results while developing the proofs. Across all 83 problems, the mean time is 61.65 minutes and the estimated mean cost is \$0.57, with a pooled cost (total cost divided by total Lean lines) of \$0.00125 per line. In a paired timing comparison with Archon \citep{frenzymath2026archon} running with Codex, on a randomly selected subset of 35 benchmark problems, Statopia's dynamic DAG procedure took 9.94\% less total elapsed time.

\paragraph{IMO 2026.}
The six IMO 2026 tasks provide a complementary test outside statistics. Unlike the statistics benchmark, these runs start from fixed Lean statements together with natural-language mathematical guidance, so they evaluate guided formal proof construction, not autonomous competition performance. The initial Lean statements are AxiomMath's compiler-compatible IMO 2026 inputs \citep{axiommath2026imo}. All six tasks were verified on their first end-to-end proving run, a first-run success rate of 100\% (6/6) in this guided setting. Table~\ref{tab:imo-results} summarizes these results and, for context, lists the times that AxiomProver reported publicly.

\begin{table}[!htbp]
\centering
\caption{Results of Statopia Prover on six guided IMO 2026 tasks. For reference, the AxiomProver times are taken from AxiomMath's public repository \citep{axiommath2026imo}. The two evaluations follow different protocols, so the timing columns are not a controlled comparison of efficiency.}
\label{tab:imo-results}
\small
\setlength{\tabcolsep}{5pt}
\begin{tabular}{@{}lrrrrr@{}}
\toprule
Task & Time (min.) & Lean lines & USD & USD/line & \shortstack{Reference \\ Time (min.)} \\
\midrule
Problem 1 & 37.63 & 720 & 0.88 & 0.00122 & 24\\
Problem 2 & 264.93 & 4,410 & 12.90 & 0.00293  & 360\\
Problem 3 & 292.68 & 7,680 & 10.89 & 0.00142 & 869\\
Problem 4 & 36.03 & 549 & 0.41 & 0.00074 & 39 \\
Problem 5 & 23.61 & 402 & 0.28 & 0.00069 & 65 \\
Problem 6 & 50.54 & 1,214 & 0.81 & 0.00067 & 139 \\
\midrule
Total & 705.42 & 14,975 & 26.17 & 0.00175 & 1,496\\
\bottomrule
\end{tabular}
\end{table}

The six Statopia runs produce 14,975 Lean lines at an estimated total off-peak cost of \$26.17, and Problems~2 and~3 account for most of the time and cost. The per-problem times add up to 705.42 minutes for Statopia and 1,496 minutes for AxiomProver. Because the two evaluations did not control for model backends, computational resources, parallelism or formal inputs, this timing comparison is only a descriptive reference. Within the reported setting, the results illustrate the practical efficiency of Statopia Prover in guided proof construction.

\subsection{Statistical Paper Developments}
\label{sec:paper-evaluation}

Each paper development starts from a target theorem, the source documents and supporting material, so it exercises material preparation, statement formalization, route assessment and proof construction together. The papers' own proofs guide these developments, but much of each formal proof still has to be discovered: published arguments often omit intermediate steps, leave regularity conditions implicit or invoke results in forms that the library does not provide, and the prover finds the missing arguments and connecting lemmas itself. Table~\ref{tab:paper-results} summarizes several archived paper cases. The file and line counts cover the complete Lean package kept for each case, including its supporting modules. The last column counts the citation inputs that each package retains, following the default described in Section~\ref{sec:engine-overview}. Five of the nine developments retain none. The other four retain between one and three each, and the results that use these inputs are verified relative to them; a user who prefers a self-contained development can ask Statopia Prover to prove these inputs as well. The rest of this subsection examines three of these cases in more detail: Gaussian approximation, CCK comparison and HMC.

\begin{table}[!htbp]
\centering
\caption{Statistical paper developments. Files and lines count the complete Lean package of each case, tasks are given as verified over total, and external citations count the cited results that each package retains as citation inputs instead of proving them (Section~\ref{sec:engine-overview}).}
\label{tab:paper-results}
\small
\setlength{\tabcolsep}{8pt}
\begin{tabular}{@{}lrrrr@{}}
\toprule
Target & Files & Lines &  \shortstack{Tasks\\ (pass/total)}& \shortstack{External \\ citations}\\
\midrule
Gaussian approximation \citep{lopes2020bootstrapping}, Thm.~3.1.  & 88 & 15,047 &6/6 & 3\\
CCK comparison \citep{chernozhukov2015comparison}, Thm.~2. & 78 & 12,223& 6/6& 1\\
Dynamic matrix recovery \citep{chen2024dynamic}, Thm.~1. & 28 & 4,827 & 2/2 & 0\\
DCF test \citep{xue2020distribution}, Thm.~1. & 164 & 41,336 & 5/5 & 1 \\
Residual spectral matching \citep{chen2024matrix}, Thm.~4.1. & 329 & 71,973 & 5/5 & 5 \\
Repeated deep regression \citep{yan2025deep}, Thm.~2.1. &  44 & 8,774  & 4/4 & 2 \\
Nonparametric prior learning \citep{jia2026nonparametric}, Thm.~2.4. & 43 & 8,159  & 1/1 & 1 \\
ProTrans \citep{chen2026deconfounding}, Thm.~1. & 88  & 11,643  & 10/10 & 0 \\
DP clustering \citep{ascolani2023clustering}, Thm.~1. & 218 & 34,932 &16/16 &0  \\
HulC \citep{kuchibhotla2024hulc}, Thm.~1. & 15 & 1,397 & 2/2 &  0\\
EILLS \citep{fan2024environment}, Thm.~4.2. & 11 & 2,079 &  1/1&0\\
HMC \citep{livingstone2019geometric}, Thm.~2.2(i--ii) and Thm.~5.14. & 73 & 7,214 &  8/8 & 1\\

\bottomrule
\end{tabular}
\end{table}

\begin{figure}[!htbp]
\centering
\resizebox{0.98\linewidth}{!}{%
\begin{tikzpicture}[
  >={Latex[length=2mm]},
  every node/.style={align=center,font=\scriptsize},
  base/.style={draw,rounded corners=1.5mm,thick,text width=27mm,minimum height=8mm,inner sep=3pt},
  verified/.style={base,draw=green!50!black,fill=green!14},
  interface/.style={base,draw=purple!65!black,dashed,fill=green!14},
  citation/.style={base,draw=purple!65!black,dashed,fill=white},
  flow/.style={->,thick,draw=gray!75!black},
  admitted/.style={->,thick,dashed,draw=purple!65!black},
  outer/.style={draw=red!65!black,fill=red!1,rounded corners=3mm,thick,inner sep=8pt}
]
\node[interface] (B) at (3.5,1.7) {\textbf{B: term II}};
\node[verified] (index) at (7,2.1) {\textbf{Index bounds}};
\node[verified] (main) at (7,0) {\textbf{Theorem 3.1\\Main}};
\node[interface] (T) at (10.5,1.7) {\textbf{T: terms I, III}};
\node[verified] (triangle) at (3.5,-1.7) {\textbf{Distance\\triangle bound}};
\node[verified] (cov) at (7,-2.1) {\textbf{Covariance\\identities}};
\node[verified] (char) at (0,1.7) {\textbf{Gaussian\\characterization}};
\node[verified] (law) at (0,0) {\textbf{Gaussian law\\identification}};
\node[verified] (orlicz) at (10.5,-1.7) {\textbf{Orlicz bound\\to integrability}};
\draw[flow] (B) -- (main);
\draw[flow] (index) -- (main);
\draw[flow] (T) -- (main);
\draw[flow] (triangle) -- (main);
\draw[flow] (cov) -- (main);
\draw[flow] (law) -- (main);
\draw[flow] (orlicz) -- (main);
\draw[flow] (char) -- (law);
\begin{scope}[on background layer]
\node[outer,fit=(B)(index)(main)(T)(triangle)(cov)(char)(law)(orlicz)] (frame) {};
\end{scope}
\node[anchor=south west,font=\bfseries\scriptsize,text=red!65!black,fill=white,inner sep=1.5pt] at ([xshift=2mm]frame.north west) {Gaussian approximation: main-task DAG};
\end{tikzpicture}%
}
\caption{The archived main-task DAG for Gaussian approximation. Its nine verified nodes comprise the main target, six local helpers and two task interfaces (purple dashed borders). B and T connect to the supporting tasks in Figure~\ref{fig:gaussian-group}. Arrows point from prerequisites to consumers.}
\label{fig:gaussian-main}
\end{figure}
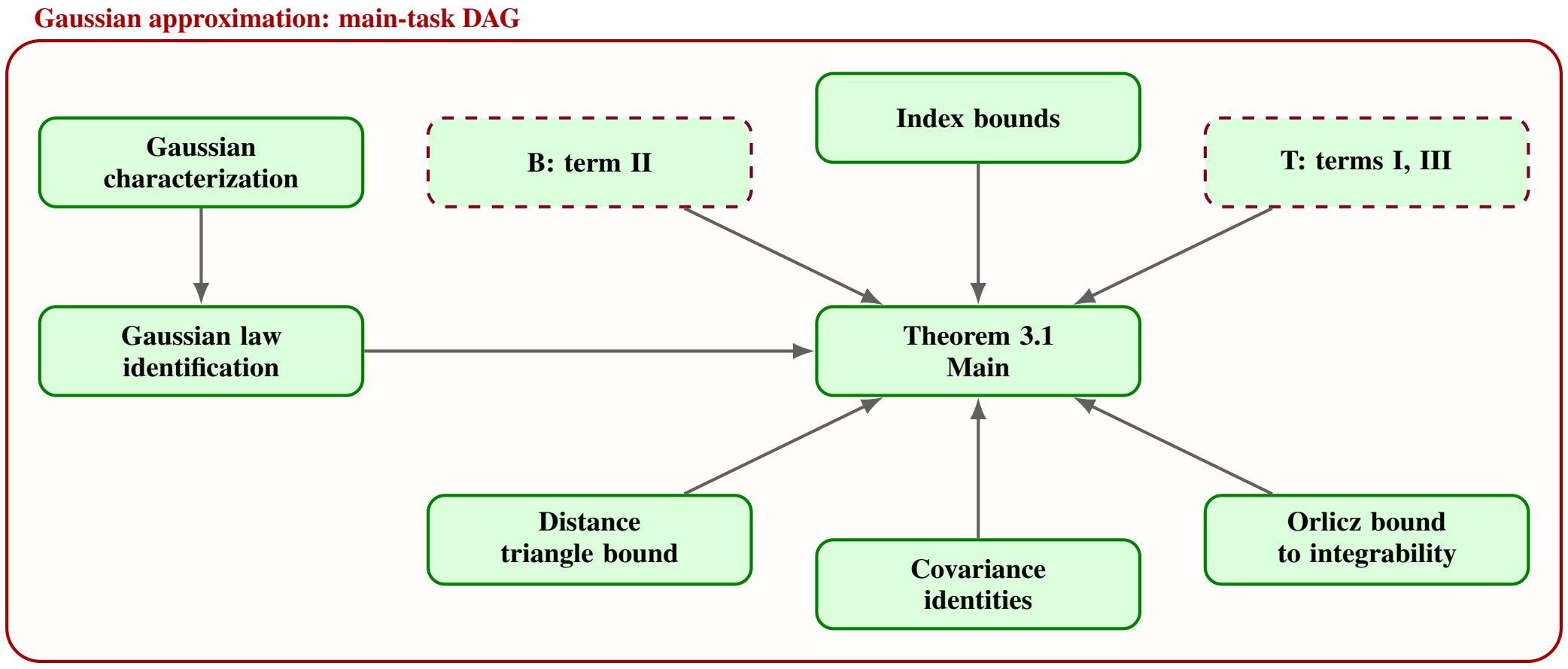

\paragraph{Gaussian approximation.}
The Gaussian approximation development proves Theorem~3.1 of \citet{lopes2020bootstrapping} (\emph{Bootstrapping Max Statistics in High Dimensions: Near-Parametric Rates Under Weak Variance Decay and Application to Functional and Multinomial Data}). Its verified proof illustrates both levels of the architecture. The main-task DAG has nine verified nodes: the main theorem, six local helpers and two interfaces supplied by supporting sub-tasks (Figure~\ref{fig:gaussian-main}). The helpers handle index bounds, the triangle inequality for the distance, covariance identities, the characterization and identification of Gaussian laws, and exponential integrability derived from an Orlicz bound.

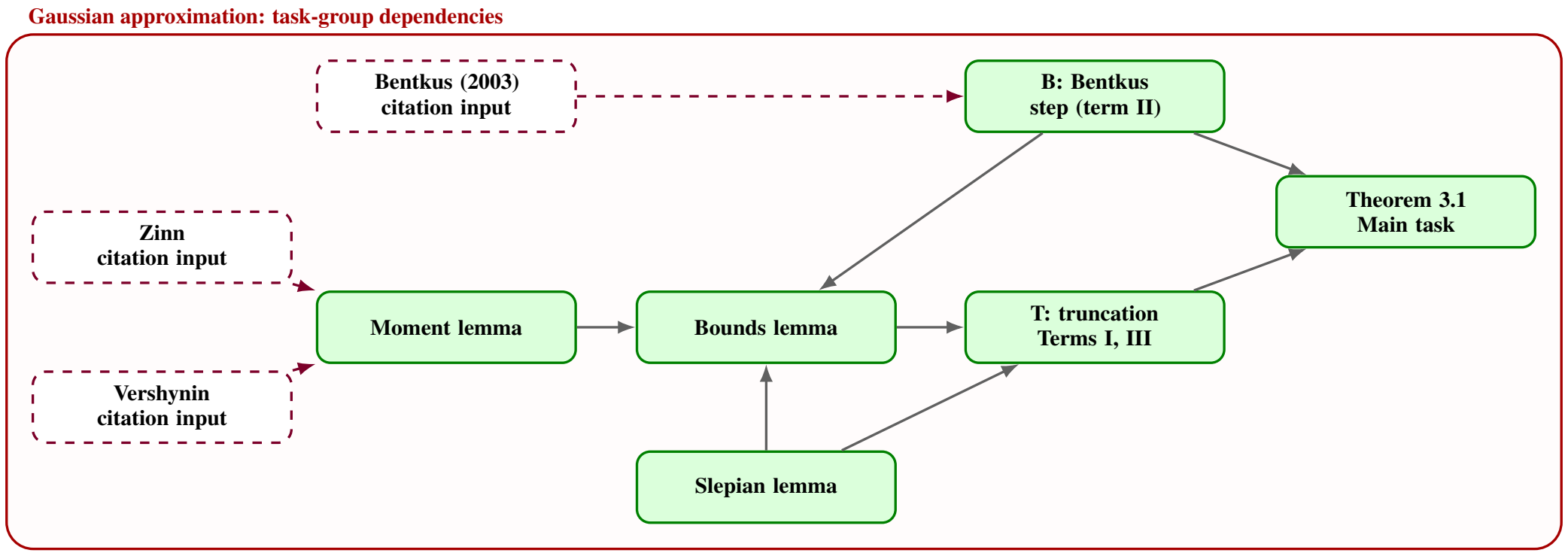
\begin{figure}[!htbp]
\centering
\resizebox{0.98\linewidth}{!}{%
\begin{tikzpicture}[
  >={Latex[length=2mm]},
  every node/.style={align=center,font=\scriptsize},
  base/.style={draw,rounded corners=1.5mm,thick,text width=27mm,minimum height=8mm,inner sep=3pt},
  verified/.style={base,draw=green!50!black,fill=green!14},
  interface/.style={base,draw=purple!65!black,dashed,fill=green!14},
  citation/.style={base,draw=purple!65!black,dashed,fill=white},
  flow/.style={->,thick,draw=gray!75!black},
  admitted/.style={->,thick,dashed,draw=purple!65!black},
  outer/.style={draw=red!65!black,fill=red!1,rounded corners=3mm,thick,inner sep=8pt}
]
\node[verified] (main) at (8,0) {\textbf{Theorem 3.1\\Main task}};
\node[verified] (B) at (4.5,1.3) {\textbf{B: Bentkus\\step (term II)}};
\node[verified] (T) at (4.5,-1.3) {\textbf{T: truncation\\Terms I, III}};
\node[verified] (bounds) at (0.8,-1.3) {\textbf{Bounds lemma}};
\node[verified] (slepian) at (0.8,-3.1) {\textbf{Slepian lemma}};
\node[verified] (moment) at (-2.8,-1.3) {\textbf{Moment lemma}};
\node[citation] (bentkus) at (-2.8,1.3) {\textbf{Bentkus (2003)\\citation input}};
\node[citation] (zinn) at (-6,-0.4) {\textbf{Zinn\\citation input}};
\node[citation] (vershynin) at (-6,-2.2) {\textbf{Vershynin\\citation input}};
\draw[flow] (B) -- (main);
\draw[flow] (T) -- (main);
\draw[admitted] (bentkus) -- (B);
\draw[flow] (bounds) -- (T);
\draw[flow] (slepian) -- (T);
\draw[flow] (B) -- (bounds);
\draw[flow] (slepian) -- (bounds);
\draw[flow] (moment) -- (bounds);
\draw[admitted] (zinn) -- (moment);
\draw[admitted] (vershynin) -- (moment);
\begin{scope}[on background layer]
\node[outer,fit=(main)(B)(T)(bounds)(slepian)(moment)(bentkus)(zinn)(vershynin)] (frame) {};
\end{scope}
\node[anchor=south west,font=\bfseries\scriptsize,text=red!65!black,fill=white,inner sep=1.5pt] at ([xshift=2mm]frame.north west) {Gaussian approximation: task-group dependencies};
\end{tikzpicture}%
}
\caption{The Gaussian task group: six verified tasks (green) and three retained citation inputs (white). Each green box represents a task with its own proof DAG. The Bentkus and Slepian tasks are reused by multiple consumers; B and T supply the interfaces in Figure~\ref{fig:gaussian-main}. Dashed arrows mark uses of citation inputs, which were kept unproved under the default setting (Section~\ref{sec:engine-overview}).}
\label{fig:gaussian-group}
\end{figure}

At the group level, five supporting sub-tasks accompany the main theorem: the Bentkus approximation step, a truncation-error comparison, a bounds lemma, a Slepian comparison lemma and a moment lemma (Figure~\ref{fig:gaussian-group}). The Bentkus step is used by both the main theorem and the bounds lemma, and the Slepian lemma is shared by the bounds and truncation tasks. Because of these shared dependencies, the group forms a DAG rather than a tree. Each task maintains its own proof development and exports a reusable interface. The integrated package retains three citation inputs, shown in white in Figure~\ref{fig:gaussian-group}. Under the default of Section~\ref{sec:engine-overview}, these cited results were not proved in this run, so the main theorem is verified relative to them.

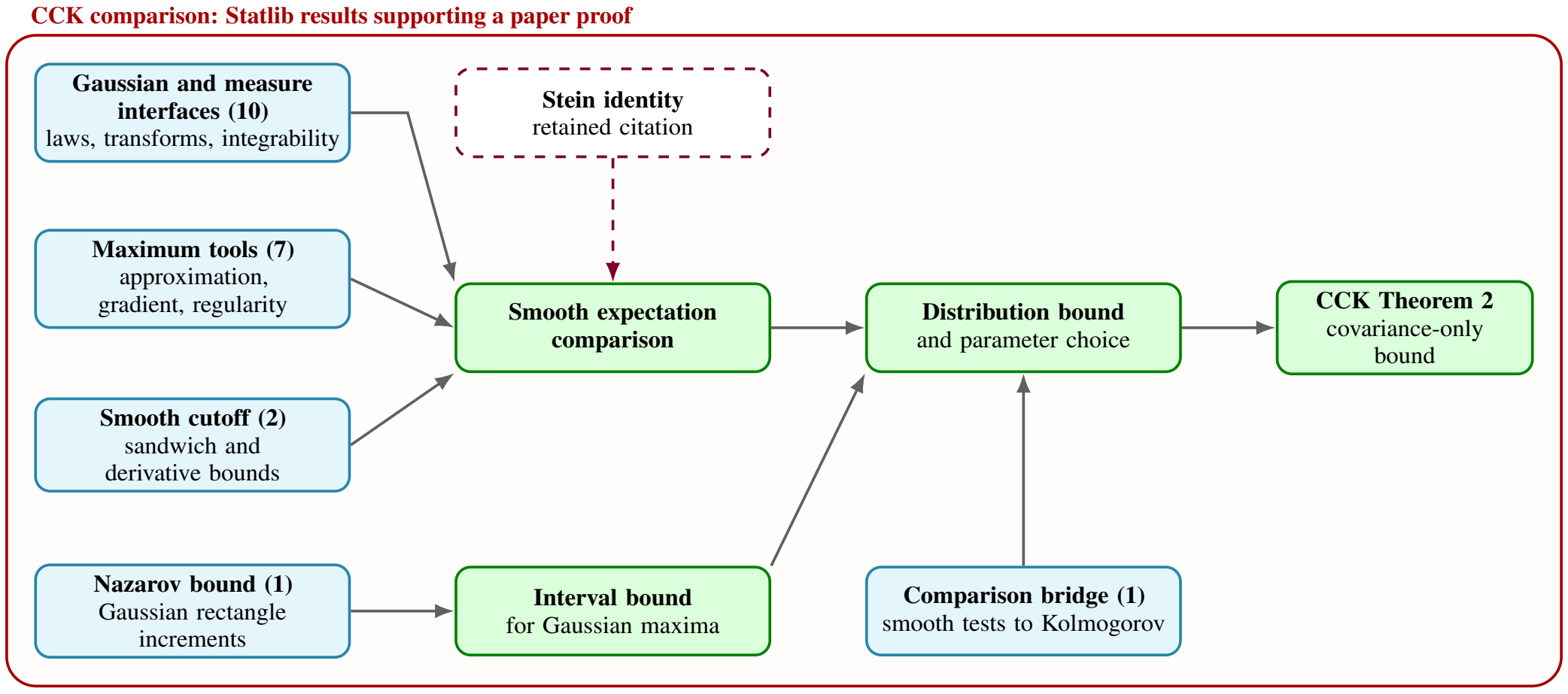
\begin{figure}[!htbp]
\centering
\resizebox{0.98\linewidth}{!}{%
\begin{tikzpicture}[
  >={Latex[length=2mm]},
  every node/.style={align=center,font=\scriptsize,
    execute at begin node={\hyphenpenalty=10000}},
  base/.style={draw,rounded corners=1.5mm,thick,
    text width=30mm,minimum height=9mm,inner sep=3pt},
  library/.style={base,draw=cyan!60!black,fill=cyan!9},
  proof/.style={base,draw=green!50!black,fill=green!14},
  citation/.style={base,draw=purple!65!black,dashed,fill=white},
  flow/.style={->,thick,draw=gray!75!black},
  admitted/.style={flow,dashed,draw=purple!65!black},
  outer/.style={draw=red!65!black,fill=red!1,
    rounded corners=3mm,thick,inner sep=8pt}
]
\node[library] (gaussian) at (0,3.6)
  {\textbf{Gaussian and measure}\\\textbf{interfaces (10)}\\laws, transforms, integrability};
\node[library] (smoothmax) at (0,1.9)
  {\textbf{Maximum tools (7)}\\approximation, gradient, regularity};
\node[library] (cutoff) at (0,0.2)
  {\textbf{Smooth cutoff (2)}\\sandwich and derivative bounds};
\node[library] (nazarov) at (0,-1.5)
  {\textbf{Nazarov bound (1)}\\Gaussian rectangle increments};

\node[citation] (stein) at (4.3,3.6)
  {\textbf{Stein identity}\\retained citation};
\node[proof] (comparison) at (4.3,1.4)
  {\textbf{Smooth expectation}\\\textbf{comparison}};
\node[proof] (anti) at (4.3,-1.5)
  {\textbf{Interval bound}\\for Gaussian maxima};

\node[library] (bridge) at (8.5,-1.5)
  {\textbf{Comparison bridge (1)}\\smooth tests to Kolmogorov};
\node[proof] (distribution) at (8.5,1.4)
  {\textbf{Distribution bound}\\and parameter choice};
\node[proof,text width=24mm] (main) at (12.4,1.4)
  {\textbf{CCK Theorem 2}\\covariance-only\\bound};

\draw[flow] (gaussian.east) -- (2.2,3.6) -- (comparison.north west);
\draw[flow] (smoothmax.east) -- (comparison.west);
\draw[flow] (cutoff.east) -- (comparison.south west);
\draw[admitted] (stein.south) -- (comparison.north);
\draw[flow] (nazarov.east) -- (anti.west);
\draw[flow] (comparison.east) -- (distribution.west);
\draw[flow] (anti.north east) -- (distribution.south west);
\draw[flow] (bridge.north) -- (distribution.south);
\draw[flow] (distribution.east) -- (main.west);

\begin{scope}[on background layer]
  \node[outer,fit=(gaussian)(smoothmax)(cutoff)(nazarov)(stein)
    (comparison)(anti)(bridge)(distribution)(main)] (frame) {};
\end{scope}
\node[anchor=south west,font=\bfseries\scriptsize,text=red!65!black,
  fill=white,inner sep=1.5pt]
  at ([xshift=2mm]frame.north west) {CCK comparison: Statlib results supporting a paper proof};
\end{tikzpicture}%
}
\caption{Statlib contributions to the final CCK comparison proof. Cyan boxes group the 21 reused named Statlib theorems, with the number of distinct theorems in parentheses; green boxes summarize their roles in the paper-specific proof. Arrows show the main mathematical uses and do not form a complete dependency graph of tasks or nodes. The interval bound is obtained from Statlib's Nazarov theorem, and the Stein identity remains an explicit citation input.}
\label{fig:cck-statlib-reuse}
\end{figure}

\paragraph{CCK comparison.}
The CCK development proves Theorem~2 of \citet{chernozhukov2015comparison} and shows how Statlib supports the mathematical content of a paper proof. The proof of its final covariance-only bound reuses 21 distinct named Statlib theorems from eight modules, and the six archived tasks of this development together reuse 29 such theorems from 12 modules. We count the distinct named theorem declarations referenced by the compiled proofs and their local helpers, excluding definitions, automatically generated proof auxiliaries and the further dependencies inside Statlib itself. Figure~\ref{fig:cck-statlib-reuse} groups these results by their role in the proof.

Gaussian and measure-theoretic interfaces provide results on linear transformations, identification of distributions and integrability. Lemmas about the maximum supply a smooth approximation of the maximum and gradient identities, together with the regularity of the coordinate maximum. Smooth cutoff bounds support the passage from smooth test functions to distribution functions. In particular, \texttt{smooth\_comparison\_}\allowbreak\texttt{to\_kolmogorov} combines a smooth comparison estimate with interval anti-concentration. The development then chooses the smoothing parameters to obtain the stated covariance-dependent rate.

For anti-concentration, Statlib supplies \texttt{gaussianVector\_}\allowbreak\texttt{rectangle\_}\allowbreak\texttt{increment\_le}, a Nazarov bound for Gaussian rectangle increments. The prover builds a local bridge from this result to an interval bound for the maximum $M_Y=\max_j Y_j$ of a centered Gaussian vector $Y=(Y_1,\dots,Y_p)$: for $p\geq2$ and $\underline{\sigma}>0$ with $\operatorname{Var}(Y_j)\geq\underline{\sigma}^{2}$ for every coordinate,
\[
  \Pr\!\left(t<M_Y\leq t+\varepsilon\right)
  \leq
  \frac{\sqrt{2\log p}+2}{\underline{\sigma}}\,\varepsilon,
  \qquad \varepsilon>0.
\]
The final comparison proof obtains its anti-concentration bound in this way, from Statlib's Nazarov theorem through the local bridge. Its smooth comparison argument still relies on Stein's integration-by-parts identity, which remains an explicitly tracked citation input. This case shows how substantial library results can be combined with paper-specific constructions and adapted interfaces to complete a larger proof development.

\FloatBarrier 
\paragraph{HMC.}
The HMC development revisits the example from Section~\ref{sec:intro}. It covers Theorem~2.2(i--ii) and the corrected Theorem~5.14 from arXiv version~5 of \citet{livingstone2019geometric}, which includes the authors' corrigendum. KAPPA also uses this paper as a case study, with a different aim: it checks the paper's own proofs of Theorem~2.2, Corollary~2.3(ii), Theorem~5.14 and Lemmas~5.15--5.17 \citep{she2026kappa}, whereas our development aims to establish the theorems themselves with verified Lean proofs. Run on the original version without access to the corrigendum, KAPPA located the same gap and proved in Lean the exact parameter range on which the original condition fails. On the corrected version, it verified Theorem~5.14 conditionally on a proposition about rejection probabilities, standard Gaussian facts and the paper's extension to a random number of leapfrog steps, while Theorem~2.2 was left unfinished because of an encoding error in a supporting statement. In our development, the corrected Theorem~5.14, stated for a fixed number of leapfrog steps, uses no citation inputs, and Theorem~2.2(i--ii) uses one. Because the two systems differ in aims, language models and budgets, this comparison is descriptive rather than controlled.

\subsection{From a Proof Development to Statlib}
\label{sec:evaluation-statlib-feedback}

Like the rest of Statlib, the component of neural network approximation was produced with Statopia Prover: the prover formalized in Lean the approximation results that underlie the regression analysis of \citet{kohler2021fully}, and the resulting development was then integrated into the library. This case illustrates how a completed proof is organized for reuse in the library (Sections~\ref{sec:statlib-modules} and~\ref{sec:library-evolution}). The development follows the constructive approximation arguments of the paper and combines explicit network constructions, local polynomial approximation and assembly over a spatial partition. In Statlib, these parts supply both the ingredients of the final approximation theorem and intermediate results that can support other developments. All seven modules of the development lie under \path{Statlib/Nonparametric/}, and the file paths below are relative to this directory.

Organizing the development for Statlib starts with a common representation of the objects used throughout the proof. The vocabulary file \path{Vocabulary/NeuralNetwork.lean} defines \texttt{DenseLayer}, \texttt{FullyConnectedReLUNet} and \texttt{reluNetworkClass} in the namespace \path{Statlib.Nonparametric.Vocabulary}. They describe, respectively, affine layers, concrete fully connected ReLU networks and the classes of functions that such networks realize within a given architecture budget. As a result, the constructions and their approximation bounds share the same definitions of realization, depth, width and parameter count. This vocabulary also connects the concrete networks produced by the construction to the function classes used in statistical approximation statements.

Related construction results and their supporting arguments are grouped into coherent source modules. For example, \path{Approximation/ReLUNetworkConstruction.lean} contains the square and multiplication constructions together with their auxiliary proofs. The tent-map iterations and state identities used to establish the square construction are private to this file, while its public approximation theorems expose the error estimates and architecture bounds that later constructions need. Selected lemmas on local polynomials and on assembly over the partition, in \path{Approximation/NeuralNetworkAlgebra.lean}, provide further reusable interfaces. By contrast, most declarations in \path{Approximation/ShallowReLUApproximationGeometry.lean}, which handles the shifted grids and phasewise estimates of the final construction, are private. A source file can thus expose several useful results while keeping together the proof steps that establish them.

Which interfaces are made public depends on how the mathematical constructions can be combined. A multiplication approximant, for instance, comes with depth and width bounds alongside its error estimate, so a later polynomial construction can track both network size and approximation accuracy. Table~\ref{tab:neural-library-reuse} summarizes representative parts of this public interface and their roles in later proofs.

\begin{table}[!htbp]
\centering
\caption{Reusable definitions and results that the neural network approximation development exposes in Statlib. The last column gives the main Statlib source files for each component, relative to \texttt{Statlib/Nonparametric/} and without the \texttt{.lean} extension.}
\label{tab:neural-library-reuse}
\small
\setlength{\tabcolsep}{4pt}
\newcommand{\filefont}{\fontsize{8}{9.5}\selectfont}
\begin{tabularx}{\linewidth}{@{}>{\raggedright\arraybackslash}p{0.16\linewidth}>{\raggedright\arraybackslash}X>{\raggedright\arraybackslash}X>{\raggedright\arraybackslash}p{0.25\linewidth}@{}}
\toprule
Library component & Public definitions and results & Purpose of the interface & Statlib files \\
\midrule
Network vocabulary & Affine layers, network realization, network classes and parameter counts & State approximation problems using shared objects and architecture budgets. & {\filefont\path{Vocabulary/NeuralNetwork}} \\
Network algebra & Linear-combination constructions, affine precomposition and depth padding & Combine or adapt networks with explicit realization and size guarantees. & {\filefont\path{Approximation/NeuralNetworkAlgebra}, \path{Approximation/ReLUNetworkConstruction}} \\
Approximation primitives & Square and multiplication approximants; centered monomial and local Taylor approximation & Build more complex approximants while tracking errors and architecture bounds. & {\filefont\path{Approximation/ReLUNetworkConstruction}, \path{Approximation/NeuralNetworkAlgebra}} \\
Local-to-global assembly & Selected tent-weight, partition and weighted approximation bounds & Combine local error estimates into an approximation bound on the domain. & {\filefont\path{Approximation/NeuralNetworkAlgebra}} \\
Statistical interfaces & Bounds transferring a candidate's pointwise or integrated squared error to network-class approximation error & Use a concrete network construction to bound the approximation term in a statistical analysis. & {\filefont\path{Approximation/NeuralNetwork}} \\
\bottomrule
\end{tabularx}
\end{table}

The final approximation result concerns H\"older-smooth functions on the unit cube and appears in Statlib as the theorem \texttt{holder\_smooth\_}\allowbreak\texttt{shallow\_relu\_}\allowbreak\texttt{neural\_network\_}\allowbreak\texttt{approximation} in \path{Approximation/ShallowReLUApproximation.lean}. Its proof goes through a statement at a fixed resolution, the integer $M$ below, which makes explicit how the network depth and the approximation error depend on $M$. Fix a dimension $d>0$, an integer $r\geq 0$, an exponent $0<\beta\leq 1$, and nonnegative ball parameters $C$ and $B$, and write $s=r+\beta$. There exist positive integers $W_0$, $M_0$ and $D_{\mathrm{block}}$, and a constant $c_{\mathrm{State}}>0$, such that for every integer $M\geq M_0$ and every function $f$ in the specified H\"older ball, the construction provides a fully connected ReLU network with depth
\[
  L_M=(D_{\mathrm{block}}+1)M^d,
\]
width at most $W_0$, and parameter count bounded by the fully connected architecture budget for $(d,L_M,W_0)$. Its realization $g$ satisfies
\[
  \lvert g(x)-f(x)\rvert
  \leq c_{\mathrm{State}}M^{-2s},
  \qquad x\in[0,1]^d.
\]
In terms of this depth, the same bound reads
\[
  \lvert g(x)-f(x)\rvert
  \leq C_{\mathrm{rate}}(L_MW_0^2)^{-2s/d},
  \qquad x\in[0,1]^d,
\]
where $C_{\mathrm{rate}}=c_{\mathrm{State}}((D_{\mathrm{block}}+1)W_0^2)^{2s/d}$. The architecture constants and both error constants do not depend on $M$, nor on $f$ within the specified ball. The public Statlib theorem states the bound in this second form, with the approximant taken from \texttt{reluNetworkClass} for the corresponding depth, width and parameter budget. The statement at a fixed resolution is private to the module. The public repository named at the start of this section exposes it as \texttt{established\_holder\_smooth\_}\allowbreak\texttt{relu\_at\_resolution} in the wrapper file \path{AtResolution.lean}, which lies outside Statlib in the folder \path{statlib-contributions/neural-network-approximation/}.

Separate bridge lemmas in \path{Approximation/NeuralNetwork.lean} connect an admissible network candidate to the approximation error of its network class: under the stated measurability, measure and class assumptions, a pointwise or integrated squared-error bound for the candidate yields a bound on the approximation error of the class. With these interfaces, later statistical proofs can use an approximation guarantee without reopening the network construction. They supply reusable ingredients for regression analysis, while estimation and risk bounds still need their own arguments.

Altogether, the development occupies seven Statlib modules with 55,922 physical lines: the vocabulary file \path{Vocabulary/NeuralNetwork.lean} and six files in \path{Approximation/}. Besides the files discussed above, these include \path{Approximation/ReLUStateMachine.lean}, which contains the state-machine constructions used for Taylor coefficients and grid scans. The line count covers the whole curated library development, including its shared definitions, public results and internal supporting proofs. The public repository also archives copies of the seven modules.

\section{Conclusion and Future Directions}
\label{sec:conclusion}

Statopia connects the construction of a formal statistical library with its use in automated formalization and proving. Statlib supplies shared statistical vocabulary, reusable interfaces and verified mathematical results, while Statopia Prover uses this infrastructure to prepare source materials, formalize statements, develop mathematical arguments, assess proof routes and construct Lean proofs. It supports formal proof discovery as well as the checking of existing proofs, and can formally verify a statement that comes without a natural-language proof. The library is itself developed with the prover, so verified proof developments can extend the infrastructure that later proofs rely on. Dynamic proof DAGs organize work within a task, and task groups connect separately developed supporting results through explicit interfaces. This organization supports parallel proof development, local revision and recovery while preserving the dependencies needed to assemble the final proof.

Several limitations remain. Formalization and proof construction can still require substantial time, particularly when statistical arguments involve missing library results, difficult representation choices or additional measurability and integrability lemmas. Repeated retrieval, interface adaptation and iterative proof repair can further prolong execution. Some tasks may remain unresolved because the required infrastructure is unavailable or an effective proof route cannot be found, even with decomposition and reuse. Such cases may require expert clarification, additional mathematical arguments or further library development. 

Future work will focus on three directions. First, we will optimize the formalization and proving workflow to reduce execution time and improve the system's ability to complete difficult tasks. Priorities include better proof route selection, less redundant retrieval and repair, and more effective coordination of parallel work. Second, we will expand Statlib's coverage of statistics and machine learning, adding both foundational infrastructure and reusable results from new formalization projects, while maintaining interfaces that support automated retrieval and composition. Third, we will extend Statopia platform to support large-scale server and cluster deployments with distributed execution, task scheduling and resource management. This will allow proof tasks to be coordinated across shared computing resources, supporting concurrent users and larger formalization workloads.

\bibliography{refer}
\bibliographystyle{icml_2025}

\end{document}